\documentclass[aps,pre,twocolumn,floatfix,superscriptaddress]{revtex4-2}
\usepackage[utf8]{inputenc}
\usepackage{mathtools}
\usepackage{amsfonts, amsmath, amsthm, amssymb, mathrsfs}
\usepackage{microtype}
\usepackage{graphicx}        
\usepackage[caption=false]{subfig}
\usepackage{booktabs}
\usepackage[usenames,dvipsnames]{xcolor}     
\usepackage{enumitem}                
\usepackage{comment}
\usepackage{hyperref}
\usepackage[capitalize]{cleveref}

\crefname{figure}{Fig.}{Figs.}
\crefname{equation}{Eq.}{Eqs.}

\hypersetup{hidelinks}

\newcommand{\figref}[1]{Fig.~\ref{#1}}

\newcommand{\vect}[1]{\boldsymbol{\mathbf{#1}}}
\newcommand{\sign}{\operatorname{sign}}
\newcommand{\pdr}[2]{\frac{\partial {#1}}{\partial {#2}}}    

\newcommand{\intz}{ \int_{-h/2}^{h/2} } 

\newcommand{\imtable}[1]{\raisebox{-.5\height}{\includegraphics[width=25mm]{images/plate-strain-stress/#1}}}

\begin{document}

\title{Odd elasticity and topological waves in active surfaces}
\author{Michele Fossati}
\affiliation{SISSA,  Trieste 34136, Italy}

\author{Colin Scheibner}
\affiliation{James Franck Institute, The University of Chicago, Chicago, Illinois 60637, USA}
\affiliation{Department of Physics, The University of Chicago, Chicago, Illinois 60637, USA}

\author{Michel Fruchart}
\affiliation{James Franck Institute, The University of Chicago, Chicago, Illinois 60637, USA}
\affiliation{Department of Physics, The University of Chicago, Chicago, Illinois 60637, USA}

\author{Vincenzo Vitelli}
\email{vitelli@uchicago.edu}
\affiliation{James Franck Institute, The University of Chicago, Chicago, Illinois 60637, USA}
\affiliation{Department of Physics, The University of Chicago, Chicago, Illinois 60637, USA}
\affiliation{Kadanoff Center for Theoretical Physics, The University of Chicago, Chicago, Illinois 60637, USA}

\date{\today}

\begin{abstract}
Odd elasticity encompasses active elastic systems whose stress-strain relationship is not compatible with a potential energy. 
As the requirement of energy conservation is lifted from linear elasticity, new anti-symmetric (odd) components appear in the elastic tensor. In this work, we study the odd elasticity and non-Hermitian wave dynamics of active surfaces, specifically plates of moderate thickness. We find that a free-standing moderately thick, isotropic plate can exhibit two odd-elastic moduli, both of which are related to shear deformations of the plate. These odd moduli can endow the vibrational modes of the plate with a nonzero topological invariant known as the first Chern number. Within continuum elastic theory, we show that the Chern number is related to the presence of unidirectional shearing waves that are hosted at the plate's boundary. 
We show that the existence of these chiral edge waves hinges on a distinctive two-step mechanism: the finite thickness of the sample gaps the shear modes and the odd elasticity endows them with chirality.  
\end{abstract}

\maketitle

\section{Introduction}

The elasticity of surfaces such as plates and shells plays an important role from biological systems~\cite{Sujit2012Delayed,Paulose2012Fluctuating} to engineered structures~\cite{reddy2006theory,Landau7,audoly2010elasticity}. 
When modeling surfaces in active and living systems, such as virus capsids~\cite{lidmar2003virus,Buenemann2007Mechanical,Douglas2001bacteriophage} or cell membranes~\cite{phillips2009physical,Seung1988defects}, active forces are often added directly into the equations of motion, while the elasticity itself is left unchanged~\cite{Needleman2017}. 
Yet, the very relationship between stress and strain can also be modified by internal energy sources.
In this situation, the elastic response can include odd-elastic coefficients~\cite{Scheibner2020,Salbreux2017,oddreview}, which describe the part of the elastic response due to non-conservative forces and therefore violate Maxwell-Betti reciprocity~\cite{Truesdell1963Meaning}.
Odd elasticity has been engineered into robotic metamaterials~\cite{Chen2021Realization,brandenbourger2021active}, and signatures have been reported in collections of spinning magnetic colloids~\cite{bililign2021chiral}, starfish embroys~\cite{tan2021development}, and models of muscular hydraulics~\cite{Shankar2022active}.
In Ref.~\cite{Salbreux2017}, the existence of these odd-elastic moduli in thin active membranes has been predicted on the grounds of symmetry.

Here we investigate the odd elasticity of thin plates and how it affects their vibrational dynamics,  focusing on the moderately thick regime in which the tilting of the plate cross-section is independent from its midplane deformation. First, a symmetry analysis reveals that an isotropic, free-standing, moderately thick plate can exhibit two independent odd-elastic moduli. 
By analyzing the normal modes of vibration of odd-elastic plates, we show that they can support edge modes in which waves propagate in a unidirectional fashion at the border of the plate. The waves propagating in these edge modes do not backscatter when they encounter sharp edges or defects. Since this robustness originates in the topological nature of the edge modes, our findings may suggest strategies for designing desirable acoustic structures, such as unidirectional waveguides~\cite{shankar2020topological,scheibner2020non,Kane2014, Bertoldi2017, huber2016topological,huber2016topological,Nash2015,Wang2015,Susstrunk2015, Foher2018Spiral,zhao2020gyroscopic,benzoni2020,Sun2020Continuum,Saremi2020Topological,shankar2021geometric}.

\section{Theory}
\subsection{Odd elasticity}
We start by reviewing the theory of odd elasticity, which is the linear elasticity of solids that exert non-conservative forces. Linear elasticity is the continuum theory that describes the behaviour of solids under small long-wavelength deformations. The deformation of the solid is described by the displacement field $\xi_i(\mathbf x) = x'_i - x_i$ giving the difference between the original position of a point $\mathbf x$ of the elastic solid and its current position $\mathbf x'$. We assume that only the variation of the internal relative distances modifies the physical state. The internal forces then depend only on the strain tensor $u_{ij} = 1/2(\partial_i \xi_j + \partial_j \xi_i)$ at linear order. 
The forces between parcels of the elastic continuum are described by the stress tensor $\sigma_{i j}$. In linear elasticity, one assumes a linear relation between stress and strain
\begin{equation}
    \label{constitutivereltensor}
    \sigma_{ij} = C_{ijk\ell} u_{k\ell}.
\end{equation}
where $C_{ijk\ell}$ is the elastic tensor, assumed here to be homogeneous in space and frequency independent. Symmetry constrains the structure of $C_{ijk\ell}$. The strain tensor is symmetric by construction and, if internal torques are absent, the stress tensor is symmetric too. In this case, the elastic tensor is symmetric under the exchanges $i \leftrightarrow j$ and $k \leftrightarrow \ell$~\cite{Landau10}. 
See Refs.~\cite{Scheibner2020,Salbreux2017}  for cases in which the displacement gradient tensor and the stress tensor are not assumed to be symmetric. 

If the system is conservative, another symmetry exists. Suppose that the system undergoes a deformation $\xi_i(t)$ in time, with strains $u_{ij}(t)$. The forces are conservative if the net work done is zero for every sequence of deformation that begins and ends in the same configuration. 
The work per unit volume of an infinitesimal deformation is given by $\sigma_{ij} du_{ij}$ and the work per unit volume under a finite cycle of deformation is calculated as a line integral. We consider a closed path $\mathscr C$ in the strain space, parameterized by $u_{ij}(\lambda)$. 
Let $\sigma_{i j}(\lambda) = C_{i j k \ell} u_{k \ell}(\lambda)$ be the associated stress tensor.
By applying Stokes's theorem, we can express the work $\mathscr W $ as the surface integral
\begin{equation}
   \mathscr W 
   = \oint_{\mathscr C} \sigma_{ij} d u_{ij} 
   = \int_\mathcal S  \frac{1}{2} \pdr{\sigma_{ij}}{u_{k\ell}} du_{ij} \wedge du_{k\ell}
\end{equation}
in which $\wedge$ is the exterior product and $\mathcal S$ is a surface in strain space such that $\partial \mathcal S = \mathscr C$. 
Using the antisymmetry of the exterior product, we can see that the forces are conservative ($\mathscr W = 0$ for all $\mathscr C$) if and only if
\begin{equation}
    \pdr{\sigma_{ij}}{u_{k\ell}} = \pdr{\sigma_{k\ell}}{u_{ij}}, 
\end{equation}
or equivalently, if and only if
\begin{equation}
    \label{Csym}
    C_{ijk\ell} = C_{k\ell ij}.
\end{equation}
This property is known as Maxwell-Betti reciprocity.
A system is said to be \emph{odd-elastic} when Eq.~\eqref{Csym} does not hold, i.e. when its elastic tensor has components that are odd under exchange $ij \leftrightarrow k\ell$ \cite{Scheibner2020}.
Note that we have made no distinction between the Cauchy and Piola-Kirchoff stress tensors, because we have assumed that there is no pre-stress in the system, see Ref.~\cite{braverman2020} for details.
Isotropy sets further constraints on the elastic tensor. Odd elasticity is incompatible with spherical isotropy (i.e. invariance under all rotations in 3D) but nonetheless it is compatible with cylindrical isotropy (i.e. invariance under all rotations preserving the $z$-axis)~\cite{Scheibner2020}. Here we employ the most general cylindrically symmetric elastic tensor, see \cref{sec:app-isotropy} for a derivation.

\subsection{Surface description: thick plate} \label{sec:kinematics}
\begin{figure}
    \centering
    \includegraphics[width=0.8\columnwidth]{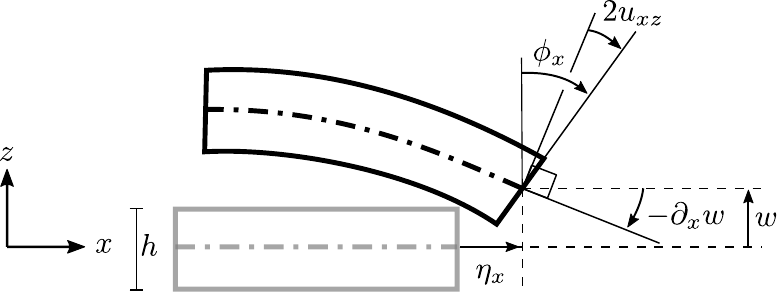}
    \caption{\textbf{Kinematics of a moderately thick plate.} 
    In the Reissner-Mindlin theory of moderately thick plates, the deformation of a plate (drawn in its undeformed reference state in grey and in a deformed state in black) is parameterized by five fields ($\eta_x$, $\eta_y$, $\phi_x$, $\phi_y$, and $w$) defined on the midplane (dash-dotted line).
    The fields $\eta_x$ and $w$ describe respectively the horizontal and vertical displacements of the midplane. The field $\phi_x$ is the angle between a deformed normal line in the $x$ direction and the $z$ axis. The quantity $-\partial_x w$ quantifies the slope of the midplane in the $x$ direction. If $-\partial_x w$ is different from $\phi_x$, then $u_{xz}$ is non-zero.  
    }
    \label{fig:plate-description}
\end{figure}

We consider a moderately thick and initially flat surface, i.e. a plate, whose midplane (the longitudinal plane that cuts the plate's thickness in half) lies at rest in the \mbox{$x$-$y$} plane, see Fig.~\ref{fig:plate-description}. The plate has uniform thickness $h$ along the $z$-axis at rest.
In the Reissner-Mindlin theory of moderately thick plates~\cite{reddy2006theory}, the full three-dimensional displacement field of the plate $\xi_i (x,y,z)$ is expressed in terms of fields defined on the horizontal midplane as
\begin{equation}
    \begin{aligned}
        \xi_x(x,y,z) &= \eta_x(x,y) + z \phi_x(x,y) \\
        \xi_y(x,y,z) &= \eta_y(x,y) + z \phi_y(x,y) \\
        \xi_z(x,y,z) &= w(x,y).
    \end{aligned}
\end{equation}
The field $\eta_{\alpha}$ represents the horizontal displacement of the midplane ($z=0$) in the direction $\alpha \in \{x,y\}$, while $w$ describes the vertical displacement. A line of points that lies vertical at rest is inclined of an angle $\phi_\alpha$ between the $z$-axis and the $\alpha$-axis after the deformation~(\figref{fig:plate-description}). While the full displacement field is defined over a three-dimensional space, the fields $\eta_\alpha$, $\phi_\alpha$ and $w$ are defined over the midplane, which is a two-dimensional manifold. 

The strains can be decomposed into two terms: $z$-independent and $z$-linear $u_{ij} = u^{0}_{ij} + z u^{1}_{ij}$. Here, $u^{1}_{ij}$ describes strains in which the top face and the bottom face of the plate are deformed oppositely. Explicitly
\begin{equation}
\label{eq:explicit-strains}
    \begin{split}
         2 u^{0}_{\alpha \beta} &=  \partial_\alpha \eta_\beta + \partial_\beta \eta_\alpha \\
        2 u^{0}_{\alpha z} &=  \phi_\alpha + \partial_\alpha w \\
        u^{0}_{zz} &= 0
    \end{split}
    \quad 
    \begin{split}
        2 u^{1}_{\alpha \beta} &= \partial_\alpha \phi_\beta + \partial_\beta \phi_\alpha  \\
        u^{1}_{\alpha \beta} &= 0 \\
        u^{1}_{zz} &= 0.
    \end{split}
\end{equation}

The bending of the midplane is quantified by $-\partial_\alpha w$, and $u^{0}_{\alpha z}$ is half of the angle between a deformed vertical line and the normal to the deformed midplane, projected in the $\alpha$ direction (\cref{fig:plate-description}). No $z$-linear term is present in the transverse strain. The vertical strain $u_{zz}$ is identically zero because the vertical displacement is independent from $z$. Since $\xi_\alpha = \eta_\alpha + z \phi_\alpha$, an originally vertical straight line remains straight after the deformation, and since $u_{zz}=0$, the line does not elongate.

\subsection{Constitutive relations}
In order to represent the tensorial constitutive relations in matrix form, we choose a basis for the strains and the stresses by defining the basis tensors
\begin{equation} \label{eq:basis}
    \begin{aligned}
        D_{ij}&=\frac{1}{\sqrt 2}
        \begin{pmatrix}
        1   &   0   &   0   \\
        0   &   1   &   0   \\
        0   &   0   &   0   \\
        \end{pmatrix}
        & S^1_{ij}&=\frac{1}{\sqrt 2}
        \begin{pmatrix}
        1  &   0   &   0   \\
        0   &   -1  &   0   \\
        0   &   0   &   0   \\
        \end{pmatrix} \\
        S^2_{ij}&=\frac{1}{\sqrt 2}
        \begin{pmatrix}
        0   &   1   &   0   \\
        1   &   0  &   0   \\
        0   &   0   &   0   \\
        \end{pmatrix}
        & T^x_{ij}&=\frac{1}{\sqrt 2} 
        \begin{pmatrix}
        0   &   0   &   1   \\
        0   &   0   &   0   \\
        1   &   0   &   0   \\
        \end{pmatrix} \\
        T^y_{ij}&=\frac{1}{\sqrt 2}
        \begin{pmatrix}
        0   &   0   &   0   \\
        0   &   0   &   1   \\
        0   &   1   &   0   \\
        \end{pmatrix}.
        & 
    \end{aligned}
\end{equation}
The $zz$ entry is zero for all the five matrices because $u_{zz}=0$ by construction and $\sigma_{zz}=0$ is further assumed (see \cref{sec:app-const-rel} for details).
Here $D$ describes cylindrical symmetric strains and stresses, $S^{1,2}$ describe the planar shears and $T^{x,y}$ the cross-section shears in $x$ and $y$ direction. This basis separates the irreducible representations of the group of rotations $SO(2)$ around the $z$ axis: $D$ is invariant, $S^\alpha$ transforms as a two-headed arrow in the plane, and $T^\alpha$ transforms as an ordinary vector in the plane.
The constitutive relation Eq.~\eqref{constitutivereltensor} can then be written in matrix form as
\begin{equation}
    \label{constitutiverelmatrix}
    \sigma_a = C_{a b} u_b
\end{equation}
where $\sigma_a$ and $u_b$ are the components of the stress and strain in the basis.

We assume the plate to be made of an homogeneous cylindrically isotropic material (see \cref{sec:app-isotropy}). The elastic tensor of the plate is obtained from the full three-dimensional elastic tensor via a reduction procedure illustrated in \cref{sec:app-const-rel}. In the basis $\{D, S^1, S^2, T^x, T^y\}$ it reads
\begin{equation}
\label{Cab_mat}
C_{ab} = 2
    \begin{pmatrix}
    \tilde{B} & 0 & 0 & 0 & 0 \\
    0 & \mu_{1} & K_{1}^{o} & 0 & 0 \\
    0 & -K_{1}^{o} & \mu_{1} & 0 & 0 \\
    0 & 0 & 0 & \mu_{2} & K_{2}^{o} \\
    0 & 0 & 0 & - K_{2}^{o} & \mu_{2}
    \end{pmatrix}
\end{equation}
with $\tilde B = \frac{-6 D^2 + 6H^2 + 9B \mu_3}{3B + 4(\sqrt 2 D + \mu_3)}$. All the moduli are inherited from the three-dimensional constitutive relations. Here, $\tilde B$ is the effective 2D bulk modulus, which relates plane isotropic dilations to the plane isotropic stress. $\mu_1$ and $\mu_2$ are passive shear moduli that couple respectively planar shears ($u_{S^1}, u_{S^2}$) and cross-section shears ($u_{T^x}, u_{T^y}$). The odd moduli $K^o_1, K^o_2$ build the antisymmetric part of the elastic tensor. $K^o_1$ maps $u_{S^1}$ to $-\sigma_{S^2}$ and $u_{S^2}$ to $\sigma_{S^1}$. $K^o_2$ does the same on the basis elements  $T^x, T^y$.

The dynamical quantities that are relevant for the plate's dynamics are the \emph{net stress tensor} $N_{ij}$ and the \emph{moment tensor} $M_{ij}$ defined by
\begin{equation} \label{eq:net-stress-moment-def}
    \begin{gathered}
        N_{ij} = \intz dz \, \sigma_{ij} \\
        M_{ij} = \intz dz\, z \sigma_{ij}.
    \end{gathered}
\end{equation}
These are respectively the zeroth and first moment of the stress in $z$. Integration in the \mbox{$z$-direction} produces a net stress that depends only on $u^0$ and a moment tensor that depends on the bending terms $u^{1}$. Using the constitutive equations Eq.~\eqref{constitutivereltensor} with the elastic tensor in Eq.~\eqref{Cab_mat}, we find that the in-plane stresses are governed by
\begin{align}
    \begin{pmatrix}
        N_D \\
        N_{S^1} \\
        N_{S^2} \\
    \end{pmatrix}&= 2 h
    \begin{pmatrix}
        \tilde B & 0 & 0 \\
        0 & \mu_1 & K^o_1 \\
        0 & -K^o_1 & \mu_1 
    \end{pmatrix}
    \begin{pmatrix}
        u^0_D \\
        u^0_{S^1} \\
        u^0_{S^2} \\
    \end{pmatrix} \label{eq:moduli1}
\end{align} 
while the bending moments are governed by

\begin{align} 
    \begin{pmatrix}
        M_D \\
        M_{S^1} \\
        M_{S^2} \\
    \end{pmatrix}
    &=  \frac{h^3}{6}
    \begin{pmatrix}
        \tilde B & 0 & 0 \\
        0 & \mu_1 & K^o_1 \\
        0 & -K^o_1 & \mu_1 
    \end{pmatrix}
    \begin{pmatrix}
        u^1_{D} \\
        u^1_{S^1} \\
        u^1_{S^2} \\
    \end{pmatrix} \label{eq:moduli2}
\end{align}
and the cross-sectional stresses by
\begin{align} 
    \begin{pmatrix}
        N_{T^x} \\
        N_{T^y}
    \end{pmatrix}
    &= 2 h
    \begin{pmatrix}
    \mu_2 & K^o_2 \\
    -K^o_2 & \mu_2 
    \end{pmatrix}
    \begin{pmatrix}
        u^0_{T^x} \\
        u^0_{T^y} \\
    \end{pmatrix} . \label{eq:moduli3}
\end{align}  

A visual representation of the components of the strain, net stress and moment tensor in the basis of Eq.~\eqref{eq:basis}, is given in \cref{tbl:plate-stress-strain}. We note that the constitutive equations in \cref{eq:moduli1,eq:moduli2,eq:moduli3} assume that the plate arises as the thin limit of a homogeneous three-dimensional solid. However, if the plate is not homogeneous along its thickness, additional moduli can appear that couple the independent equations in \cref{eq:moduli1,eq:moduli2,eq:moduli3}, see \cref{sec:app-const-rel}. Notice that the moduli $\mu_2$ and $K^o_2$ set the stresses in response to the cross section shearing. The matrix in Eq.~(\ref{eq:moduli3}) is proportional to a rotation matrix whose chirality is set by the modulus $K^o_2$. This will play a crucial role when we discuss the specturm and topological modes in Sections~\ref{sec:topologicalwaves}.

\begin{table*}[ht]
    \centering
   \begin{tabular}{lcccccc}
       \toprule
         & $D$ & $S^1$ & $S^2$ & $T^x$ & $T^y$ \\
        \midrule
        $u^0$ & \imtable{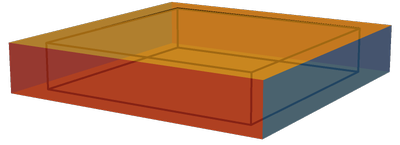} & \imtable{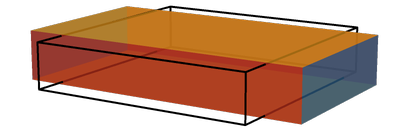} & \imtable{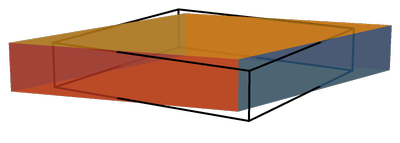} & \imtable{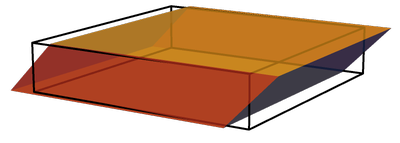} & \imtable{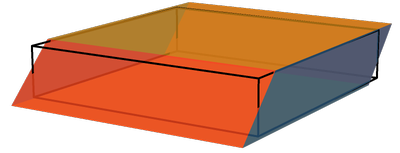}\\
        $u^1$ & \imtable{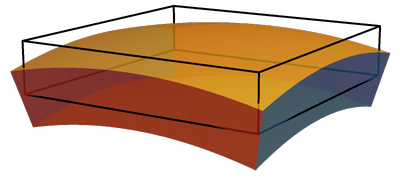} & \imtable{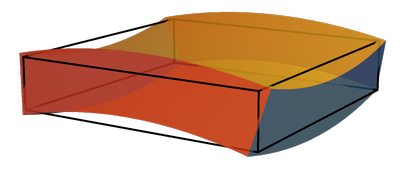} & \imtable{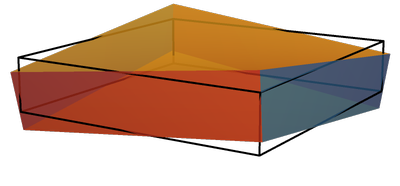} &  & \\
        $N$ & \imtable{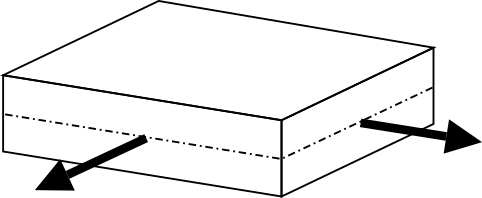} & \imtable{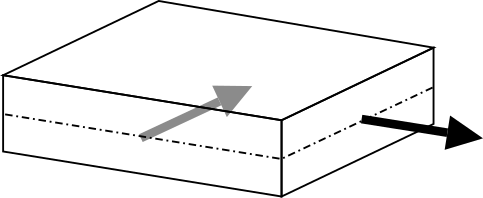} & \imtable{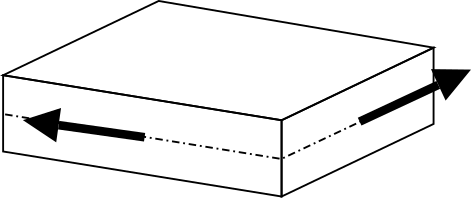} & \imtable{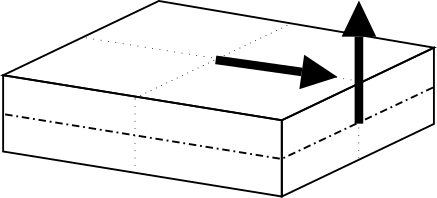} & \imtable{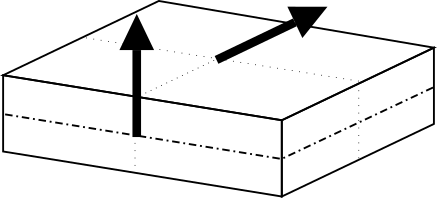}\\
        $M$ & \imtable{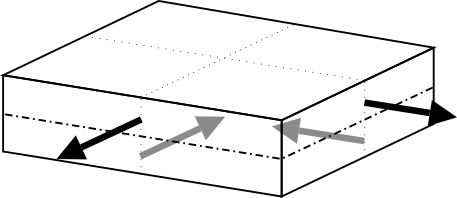} & \imtable{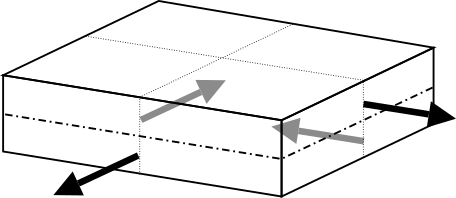} & \imtable{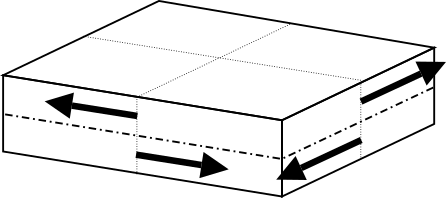} &  & \\
        \bottomrule
    \end{tabular}
    \caption{  
    \textbf{Plate strains and stresses.}
    Here, $u^{0}$ and $u^{1}$ are respectively the zeroth and first order contributions to the strain in a power expansion in $z$; $N$ is the net stress and $M$ is the bending moment. 
    The column is labeled by the tensor basis element on which the stresses, strains, and moments are projected. 
    The arrows are visual cues for the forces and torques on each of the shown surfaces (Light arrows are pointing towards the inside of the plate).  } 
    \label{tbl:plate-stress-strain}
\end{table*}

Having the constitutive relations, we can examine the linearly independent cycles in strain space over which work is extracted $\mathscr C = \partial \mathcal S$. The work per unit surface is
\begin{equation}
\begin{split}
    \mathcal W 
    &= \intz dz \int_{\mathscr C} du_a \, C_{ab} u_b  \\
    &= h \int_{\mathscr C} du^0_a \, C_{ab} u^0_b  + \frac{h^3}{12} \int_{\mathscr C} du^{1}_a \, C_{ab} u^{1}_b \\
    &= \frac{h}{2} \int_\mathcal S C_{ab} \, du^{0}_a \wedge du^{0}_b + \frac{h^3}{24} \int_\mathcal S C_{ab} \, du^{1}_a \wedge du^{1}_b.
\end{split}
\end{equation}

\begin{figure}
    \begin{center}
        \subfloat[][Planar shearing]
           {\includegraphics[width=0.48 \columnwidth]{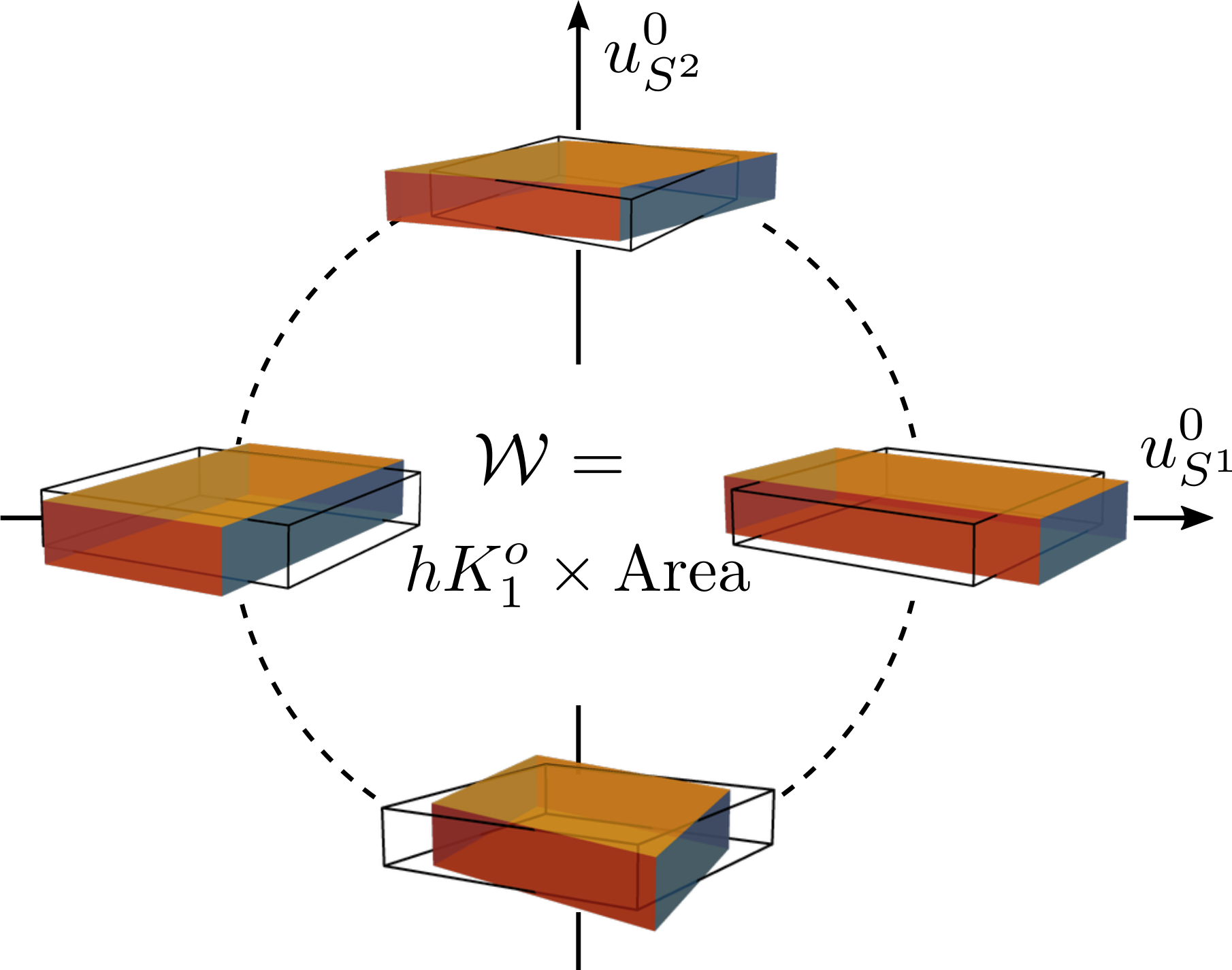}}
        \quad
        \subfloat[a][Bending ]
           {\includegraphics[width=0.48 \columnwidth]{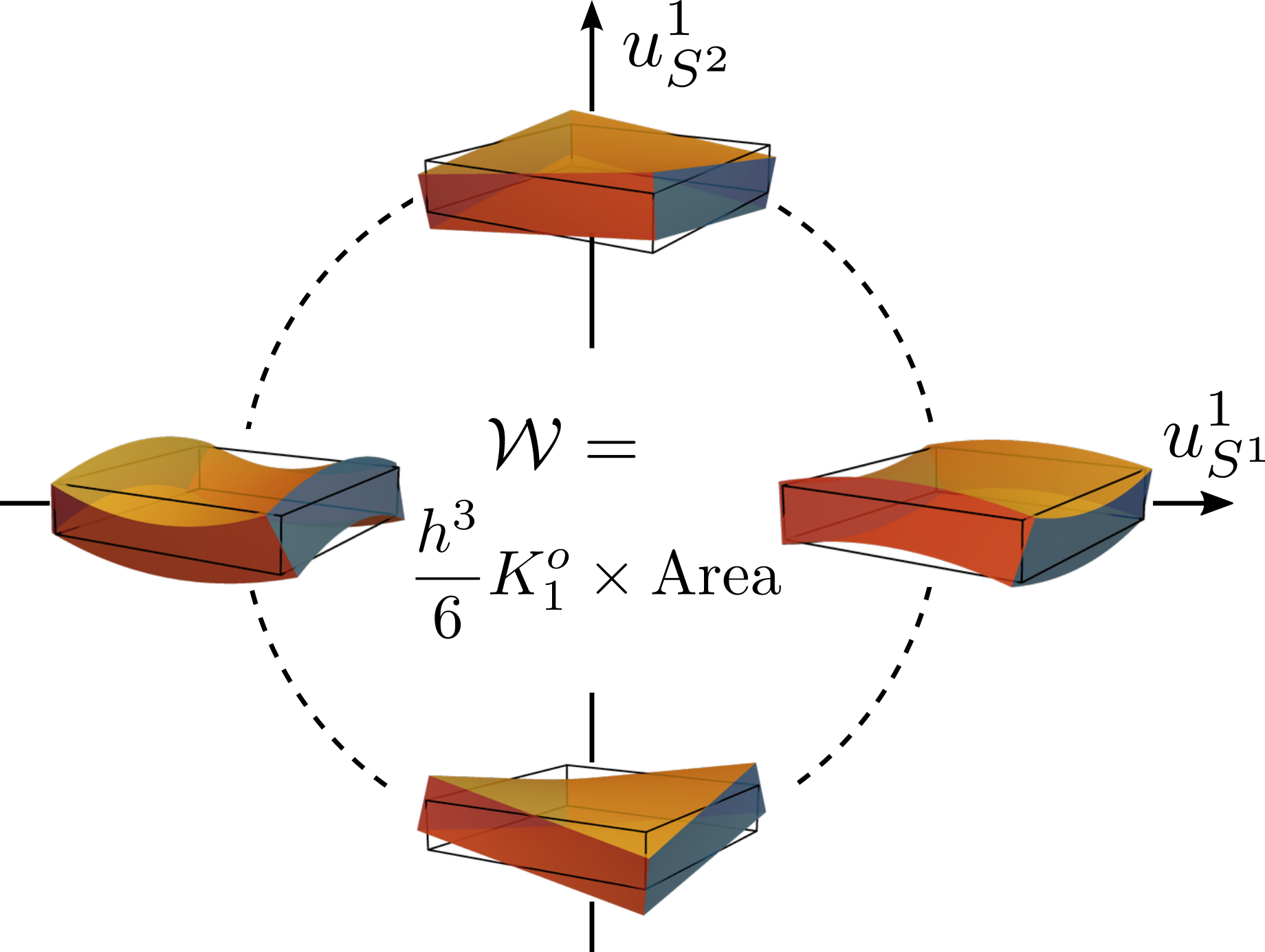}}
        \\
        \subfloat[][Cross-section shearing]
           {\includegraphics[width=0.48\columnwidth]{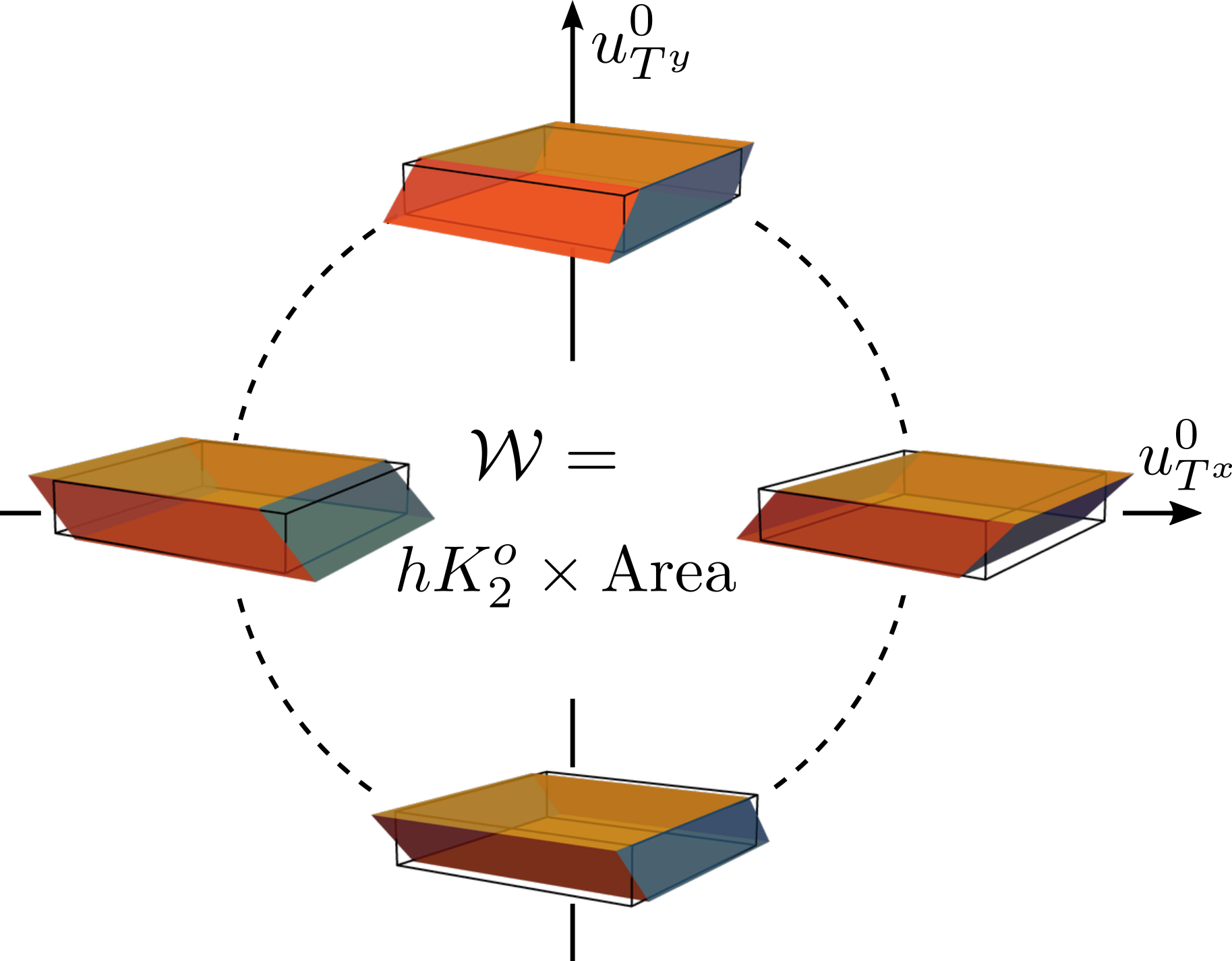}}
    \end{center}
    \caption{ \textbf{Energy cycles in the elastic plate.}
    Illustrations of the three linearly independent cycles that extract energy from the plate. Panel (a) utilizes in planar shear deformations, Panel (b) utilizes bending deformations, and Panel (c) utilizes cross-section shearing. See Table~\ref{tbl:plate-stress-strain} for definitions of the deformation icons.  
    } 
    \label{fig:cycles}
\end{figure}

There are three independent ways to extract energy with a cycle of deformations, represented in Fig.~\ref{fig:cycles}. Cycling in the plane $u^0_{S^1}$-$u^0_{S^2}$, the energy density extracted is equal to $2 h K^0_1$ times the area enclosed in the strain space (\figref{fig:cycles}a). A bending cycle that involves $u^1_{S^1}$ and $u^1_{S^2}$ extracts $(h^3/6) K^o_1$ times the area enclosed (\figref{fig:cycles}b). With a cycle in the $u^0_{T^x}$-$u^0_{T^y}$ plane, the density of work is $2 h K^0_1$ times the area enclosed (\figref{fig:cycles}c).

\subsection{Equations of motion}

We now move on to the dynamics of the system. To do so, let us assume that the elastic material evolves according to a Newtonian dynamics, and is subject to an external friction force $f^{\mathrm{ext}}_i = - \Gamma \dot \xi_i$ summarizing the effect of the environment.
The evolution of the plates is then described by the following conservation laws for linear and angular momentum (derived in \cref{sec:app-virtual-work})
\begin{align}
    h( \rho \ddot\eta_\alpha + \Gamma \dot \eta_\alpha )  &= \partial_\beta N_{\beta \alpha} \label{eq:eom-eta-raw}\\
    h(\rho  \ddot w  + \Gamma \dot w) &=  \partial_\alpha N_{\alpha z} \label{eq:eom-w-raw} \\
    \frac{h^3}{12}(\rho \ddot \phi_\alpha + \Gamma \dot \phi_\alpha) &= \partial_\beta M_{\beta \alpha} - N_{z \alpha}. 
\end{align}
where $\rho$ is the density of the plate, assumed to be uniform.

Using the constitutive relations, \cref{eq:moduli1,eq:moduli2,eq:moduli3}, we obtain
\begin{gather}
    \label{eq:eom-eta}
    \rho \ddot \eta_\alpha + \Gamma \dot \eta_\alpha = (B \partial_\alpha \partial_\beta + \mu_1 \nabla^2 \delta_{\alpha\beta} + K^o_1 \nabla^2 \epsilon_{\alpha\beta}) \eta_\beta \\
    \label{eq:eom-w}
    \rho \ddot w + \Gamma \dot w = \mu_2 \nabla^2 w + \mu_2 \partial_\alpha \phi_\alpha + K^o_2 \epsilon_{\alpha\beta} \partial_\alpha \phi_\beta\\
     \begin{multlined}
        \label{eq:eom-phi}
         \rho \frac{h^2}{12} \ddot \phi_\alpha + \Gamma \dot \phi_\alpha = \frac{h^2}{12} (B \partial_\alpha \partial_\beta + \mu_1 \nabla^2 \delta_{\alpha\beta} + K^o_1 \nabla^2 \epsilon_{\alpha\beta}) \phi_\beta \\
           - (\mu_2 \delta_{\alpha\beta} + K^o_2 \epsilon_{\alpha\beta} ) (\partial_\beta w + \phi_\beta)
    \end{multlined} 
\end{gather}
The dynamics of the horizontal displacements $\eta_\alpha$ is decoupled from the other degrees of freedom. It describes a purely two dimensional odd-elastic and isotropic system, which has already been studied in~\cite{Scheibner2020}. The out-of-plane dynamics governed by \cref{eq:eom-w,eq:eom-phi} will be the focus of the following section.

\section{Normal modes of vibration and topological waves} \label{sec:topologicalwaves}
\begin{figure*}
    \begin{center}
        \subfloat[][Passive eigenmode]
           {\includegraphics[width=0.6\columnwidth]{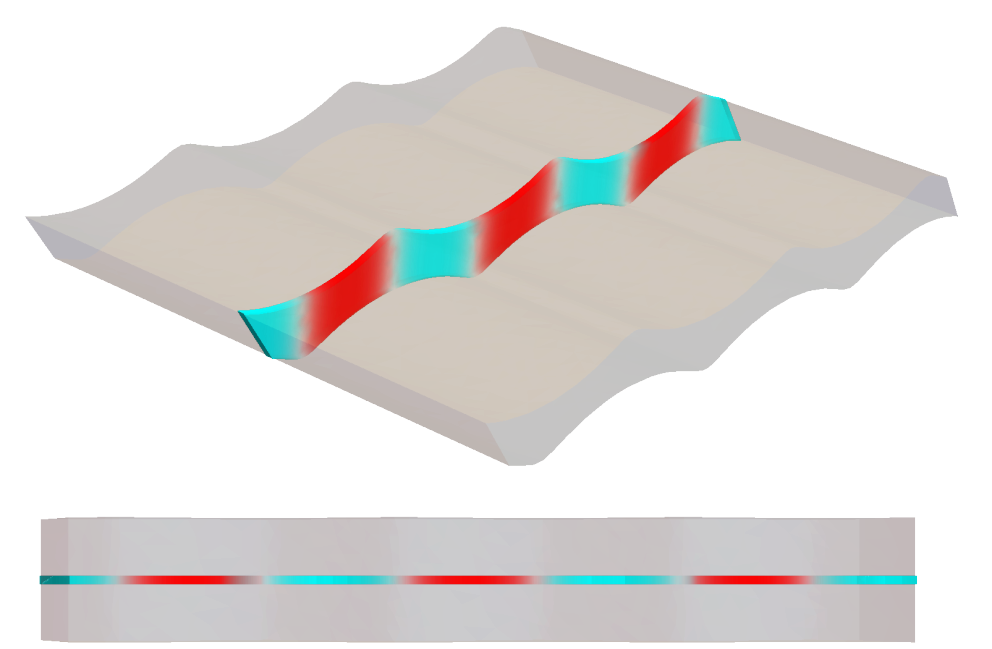}}
        \qquad
        \subfloat[][Active eigenmode ]
           {\includegraphics[width=0.6 \columnwidth]{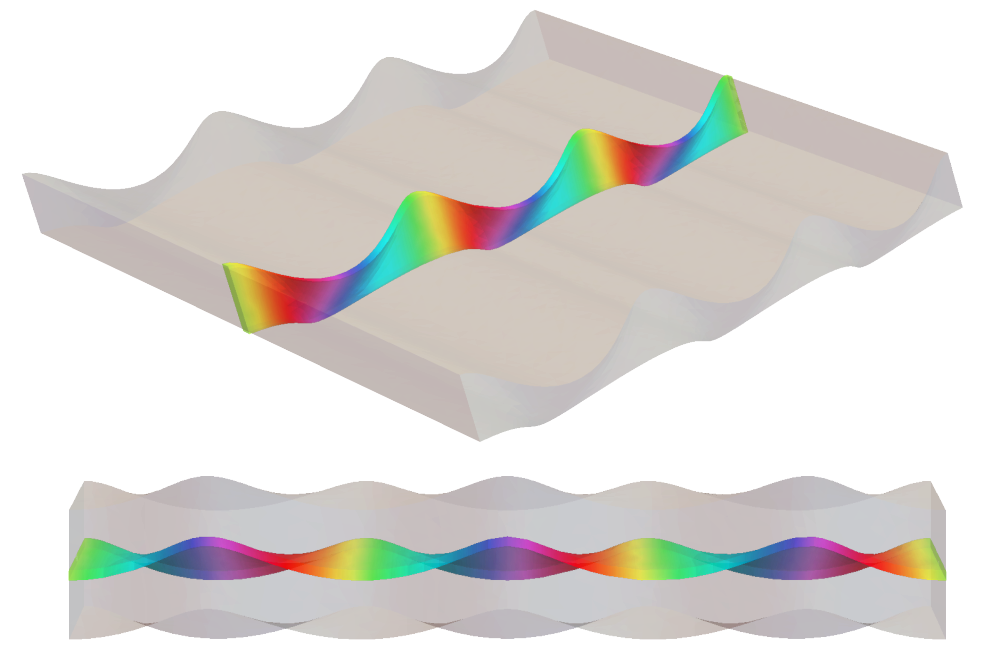}}
        \qquad
        \subfloat[][Legend: strain space]
           {\includegraphics[width=0.4\columnwidth]{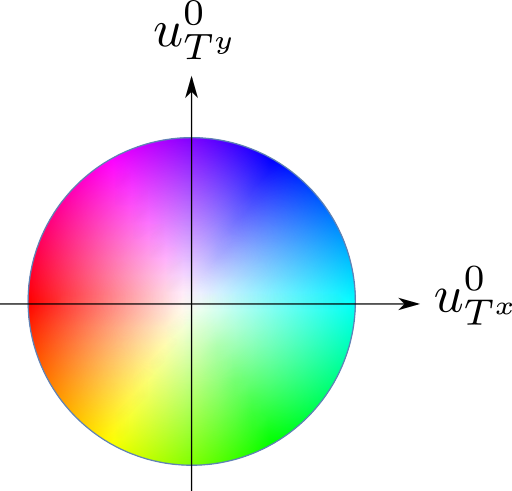}}
    \end{center}
    \caption{
    \textbf{Perspective and top view of bulk, gapped eigenmodes.} We illustrate the eigenmodes implied by Eq.~(\ref{eq:eigenvalue-problem}). As a guide, a deformed cross-sectional slice of the plate is coloured by the local component of cross-section shears $u^0_{T^{x,y}} $ (see Fig.~\ref{fig:cycles} and Table~\ref{tbl:plate-stress-strain} for illustration of cross-section shears). {\bf a.}~A passive eigenmode ($K_1^o = K_2^o=0$) traces out a line (red-blue) in strain space. {\bf b.}~An eigenmode with $K_2^o >0$ and $\tilde B= \mu=K_1^o=0$ draws an ellipse in the strain plane (full color wheel present). The energy injected in the system after a complete oscillation is proportional to the area covered in the strain plane.}
    \label{fig:bulk-eigenmode}
\end{figure*}

We now analyze the normal modes of vibration of the odd-elastic plate.
As illustrated in Fig.~\ref{fig:bulk-eigenmode}, we first show how the flexural eigenmodes behave in the bulk of the plate (i.e. far from any edge). 
Then, we show how to assign a topological invariant to the band structure of the system. This step allows us to predict the existence of unidirectional edge states exponentially localized to the boundary of the plate (Fig.~\ref{fig:edge-drawing}), under certain boundary conditions~\cite{delplace2017topological,souslov2019topological,tauber2019bulk-interface,tauber2020anomalous}.

\subsection{Normal modes of vibration}

The out of plane deformations of the plate are described by \cref{eq:eom-w,eq:eom-phi}. Since they are linear, we look for plane waves solutions of the form $\phi_\alpha (\vect x,t) = \phi_\alpha e^{i(\vect k \cdot \vect x - \omega t) }$ and $w(\vect x, t) = w e^{i(\vect k \cdot \vect x - \omega t)}$ and we  investigate the dispersion relations and the eigenmodes. Using the normalized quantities $\mathbf q = h \sqrt{12} \mathbf k$ and  $ \tilde w = \frac{w}{\sqrt{12} h} $, we obtain
\begin{equation}     \label{eq:eigenvalue-problem}
    \frac{h^2}{12} (\rho \omega^2 + i \Gamma \omega) \begin{pmatrix} \vect \phi \\ \tilde w \end{pmatrix}
    = \mathcal M (\vect q) \begin{pmatrix} \vect \phi \\ \tilde w \end{pmatrix}
\end{equation}
in which we have introduced the dynamical matrix 
\begin{widetext}
\begin{equation}
    \mathcal{M}( \vect q) = 
    \begin{pmatrix}
        \tilde B q_x^2 + \mu_1 q^2 + \mu_2 & \tilde B q_x q_y + K^o_1 q^2 + K^o_2 &  i(\mu_2 q_x + K^o_2 q_y) \\
        \tilde B q_x q_y - K^o_1 q^2 - K^o_2 & \tilde B q_y^2 + \mu_1 q^2 + \mu_2  & i(\mu_2 q_y - K^o_2 q_x) \\
        -i (\mu_2 q_x - K^o_2 q_y) & -i (\mu_2 q_y + K^o_2 q_x) & \mu_2 q^2
    \end{pmatrix}.
\end{equation}
\end{widetext}
The dynamical matrix $\mathcal M$ is non-Hermitian due to the presence of the odd moduli. The spectrum depends only on $q= |\mathbf q|$ because of the cylindrical isotropy of the constitutive relations. Since the characteristic polynomial of the matrix is real, the eigenvalues are either all real, or one eigenvalue is real and the two other are complex conjugate of each other. The spectrum contains in three bands in the complex plane, symmetric under reflection with respect to the real axis. 

When $\vect q =0$ the matrix becomes
\begin{align}
    \mathcal{M}(\vect q = 0 ) = \begin{pmatrix}
    \mu_2 & K_2^o & 0 \\
    - K_2^o & \mu_2 & 0 \\
    0 & 0 & 0 
    \end{pmatrix}.
\end{align}
As the eigenvalues $\lambda = \mu_2 \pm i K_2^o$ of $\mathcal{M}(\vect q =0) \neq 0$ are non-zero, the system has a gap in its dispersion relation. The origin of this band gap can be traced to the small but finite thickness of the beam. Recall that plane waves in a three-dimensional solid are described by a wave vector with three components $(q_x, q_y, q_z)$. Since the plates we consider are thin the $z$ direction, the minimum nonzero value of $q_z$ is proportional to $1/h$, resulting finite frequency modes even when $q_x=q_y=0$. This gap is proportional to the moduli $\mu_2$ and $K_2^o$ because these moduli respond to cross section shearing, which involves gradients in the $z$ direction (cf. Eq.~(\ref{eq:moduli3}) and Table~\ref{tbl:plate-stress-strain}).    

\begin{figure*}
    \centering
    \includegraphics[width=\textwidth]{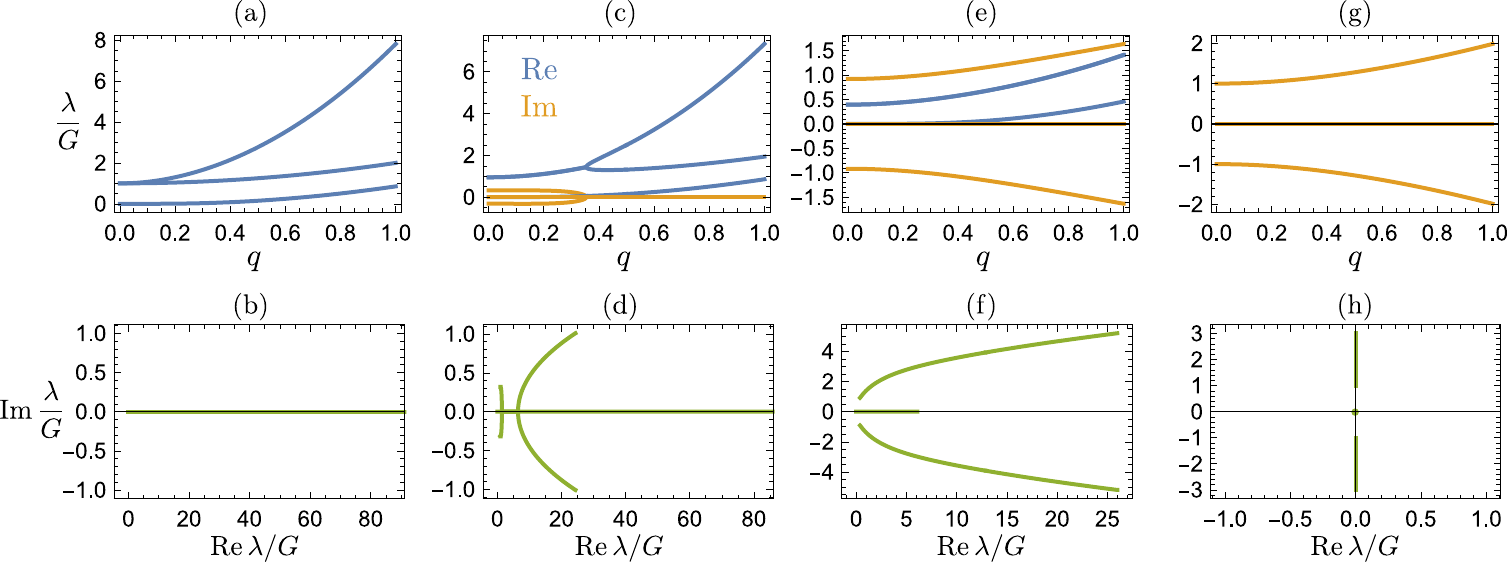}
    \caption{
    \textbf{Eigenvalues of the dynamical matrix in the complex plane.} In (a,c,e,g) we plot real and imaginary part of the spectrum, and in (b,d,f,h) we plot the spectrum parametrically in the complex plane for $q \in [0,5]$. The eigenvalues are expressed in units of $G=\sqrt{\mu_2^2 + {K^o_2}^2}$. In (a,b) the odd moduli are zero and thus the spectrum is real. $\tilde B/G = 5.7, \mu_1/G = 1, \mu_2/G = 1, K^o_1/G = 0.0, K^o_2/G = 0.0$.  In (c,d) the odd modules are turned on and the spectrum shows an imaginary part. $\tilde B/G = 5.4, \mu_1/G = 0.95, \mu_2/G = 0.95, K^o_1/G = 0.3, K^o_2/G = 0.3$. Lowering the (passive) bulk modulus and raising the odd moduli in (e,f), the spectrum becomes gapped. $\tilde B/G = 1.3, \mu_1/G = 0.4, \mu_2/G = 0.4, K^o_1/G = 0.66, K^o_2/G = 0.92$. In (g,h) the spectrum is completely imaginary as only the odd moduli are nonzero. $K^o_1/G = 0.7, K^o_2/G = 1$.}
    \label{fig:spectrum-complex}
\end{figure*}

In \cref{fig:spectrum-complex}, we plot the spectrum for different values of $\tilde B$, $\mu_1$, $\mu_2$, $K^o_1$ and $K_2^o$. When the odd moduli vanish, the dynamical matrix is Hermitian and therefore the spectrum is real (\cref{fig:spectrum-complex} a,b). 
The normal modes here contain two gapped modes whose deformation is dominated by cross-section shearing.
When the odd moduli are nonzero but sufficiently small (\cref{fig:spectrum-complex} c,d), the spectrum has a nonzero imaginary part, but the gap is still sorted by the real (blue) part of the spectrum.   
By contrast, for sufficiently large $K_2^o/\mu_2$, the gap opens in the imaginary part of the (\cref{fig:spectrum-complex} e,f). Finally, when the even moduli vanish, the spectrum is entirely imaginary (\cref{fig:spectrum-complex} g,h). When the gap is in the imaginary part, the chiral modulus $K_2^o$ dominates over $\mu_2$, and hence we may intuitively expect chiral phenomenology in the wave mechanics. In the next section, we show that when the gap is open at sufficiently large $K_2^o$, the combination of chirality and a band gap gives rise to a nonzero topological invariant known as the Chern number, that leads to unidirectional wave propagation at the boundary of a finite system.

\subsection{Topological invariant}

The bulk spectrum of $\mathcal{M}(\vect q)$ corresponds to the vibrational modes of an infinite, unbounded plate. Here, we utilize the topological properties of the eigenvectors to show that when a boundary is introduced into the plate, the boundary will host unidirectional edge states (subject to appropriate boundary conditions). 
In \cref{sec:app-chern}, we compute a topological invariant known as the first Chern number. 
This quantity can be expressed as the integral of the so-called Berry curvature over the momentum space. In most applications of topological band theory, the structure under consideration takes the form of a discrete lattice, and so the momentum space is a closed manifold (namely a torus). By contrast, here we analyze continuum equations for which the momentum space corresponds to the real plane. This introduces additional subtleties regarding meaning and interpretation of the Chern number~\cite{delplace2017topological,souslov2019topological,tauber2019bulk-interface,tauber2020anomalous,Volovik1988,So1985,Coste1989,Bal2019,Bal2019b}. 
These aspects have been studied in the context of fluids with broken microscopic time-reversal and parity symmetries exhibiting a property known as odd viscosity~\cite{Avron1998}. In \cref{app:mapping}, we provide an explicit mapping between the odd-elastic plate and these fluids. Other (distinct) topological properties that have been studied in elastic continua include topological softness associated with floppy mechanisms in the micro-structure~\cite{Sun2020Continuum,Saremi2020Topological}. 

For simplicity, in the main text, we concentrate on the simplest scenario in which the bands are gapped is the purely active plate, where the dynamical matrix
\begin{equation}
    \mathcal{M}( \vect q) = 
    \begin{pmatrix}
        0 &  K^o_1 q^2 + K^o_2 &  i K^o_2 q_y \\
       - K^o_1 q^2 - K^o_2 & 0& - i K^o_2 q_x \\
        i K^o_2 q_y & -i K^o_2 q_x & 0
    \end{pmatrix}
\end{equation}
has a flat band of eigenvalues $\lambda_0(q) = 0$ and two purely imaginary bands
\begin{equation}
    \lambda_\pm(q) = \pm i \sqrt{(K^o_2 q)^2 + (K^o_1 q^2 + K^o_2)^2}.
\end{equation}
The spectrum has a gap whenever $K^o_2 \neq 0$. As detailed in \cref{sec:app-chern}, if $K^o_1 \neq 0$ the eigenvectors of $\mathcal M (\vect q)$ do not depend of the direction of $\vect q$ in the limit $\vect q \to \infty$. This allows us to compactify the momentum space to a sphere, identifying all the points at infinity. It is then possible to associate a topological invariant to the bulk eigenvectors associated with each band, the first Chern number $\mathcal C$. Its values are (see \cref{sec:app-chern})
\begin{equation} \label{eq:chern-result}
    \begin{aligned}
    \mathcal C_0 &= 0 \\
    \mathcal C_+ &= \sign K^o_2 - \sign K^o_1\\
    \mathcal C_- &= - \mathcal C_+.
    \end{aligned}
\end{equation}

\begin{figure}
    \centering
    \includegraphics[width=0.8\columnwidth]{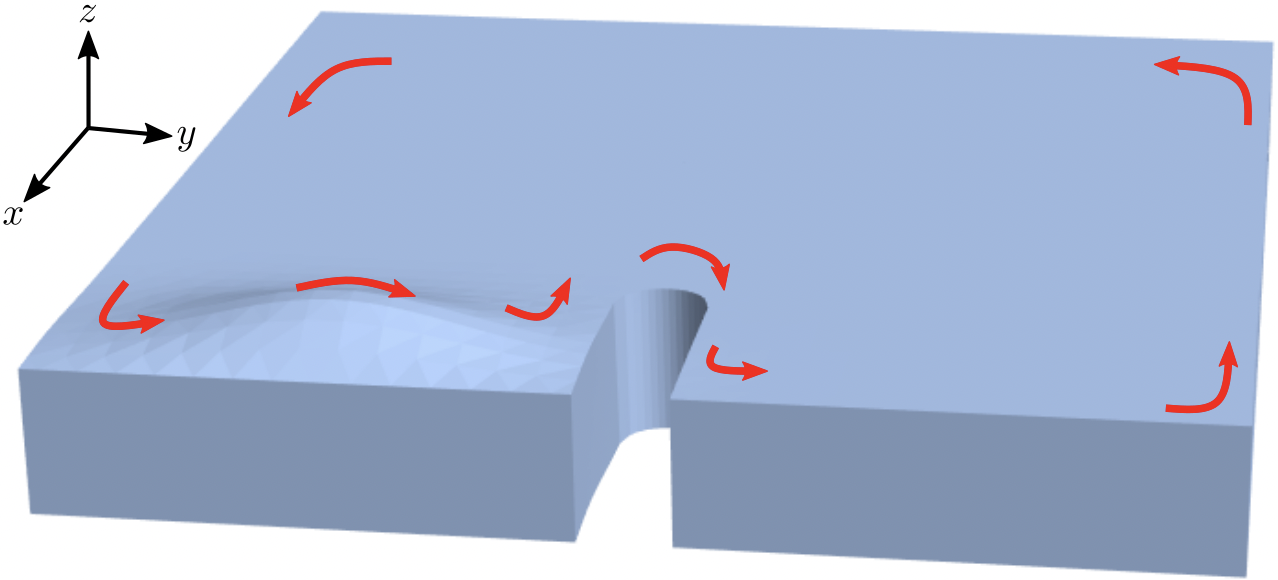}
    \caption{\textbf{Topological edge mode.}
    A schematic illustration of a topological elastic wave travelling unidirectionally along the boundary of an odd-elastic plate.
    }
    \label{fig:edge-drawing}
\end{figure}

The presence of non-zero Chern numbers suggests that edge modes could appear at the system boundary~\cite{Hasan2010}. 
In the continuum, the precise boundary conditions play a subtle role in determining the number of edge modes~\cite{tauber2019bulk-interface,tauber2020anomalous}.  
For concreteness, we now consider a finite plate that has its midplane defined over a region with boundary $\mathcal B$.
We assume that the displacement field vanishes at the boundary:
\begin{equation}
    w |_{\mathcal B} = 0  \quad \phi_x |_{\mathcal B} = 0 \quad \phi_y |_{\mathcal B} = 0.
\end{equation}
In the context of chiral active fluids, it has been verified numerically in~\cite{souslov2019topological} that under this condition the system displays topologically protected edge modes (under the mapping of \cref{app:mapping}, the boundary condition above corresponds to $\rho = 0, \vect v = 0$ at the fluid's boundary). A sketch of the edge mode in the plate is shown in Fig.~\ref{fig:edge-drawing}.

Finally, we comment that our analysis relies on the Reissner-Mindlin theory of thin plates. This theory is applicable for low frequency modes. 
Hence, our analysis is only valid when the band gap is small enough to lie withing the range of validity of the low frequency approximation. The precise range of validity often requires experimentation, but is generally larger for floppier materials with a lower shear modulus.

\section{Conclusion}
In this work, we studied the equations of motion for an odd-elastic plate in the Reissner-Mindlin moderately thick limit. The in-plane dynamics of the plate follows the equations of two-dimensional odd elasticity. The out-of-plane dynamics displays new effects. The spectrum is gapped as long as the active elastic moduli are dominant with respect to the passive ones. Besides, the bands can aquire a non-zero Chern number, and this leads to the apparition of unidirectional waves at the boundary of a finite plate.

\section{Acknowledgements} 
M.Fo. would like to thank L. Molinari for fruitful conversation.
C.~S.~acknowledges support from the Bloomenthal Fellowship and the National Science Foundation Graduate Research Fellowship under Grant No.~1746045. 
M.Fr. acknowledges support from a MRSEC-funded Kadanoff–Rice fellowship (DMR-2011854) and the Simons Foundation.
V.V.~acknowledges support from the Simons Foundation, 
the National Science Foundation under grant DMR-2118415, and the University of Chicago Materials Research Science and Engineering Center, which is funded by the National Science Foundation under Award No.~DMR-2011854. 
This research was also sponsored by the Army Research Office and was accomplished under
Cooperative Agreement Number W911NF-22-2-0109. The views and conclusions contained in this document are
those of the authors and should not be interpreted as representing the official policies, either expressed or implied, of
the Army Research Office or the U.S. Government. The U.S. Government is authorized to reproduce and distribute
reprints for Government purposes notwithstanding any copyright notation herein.

\bibliographystyle{naturemag}
\bibliography{odd-plates}

\begin{thebibliography}{10}
\expandafter\ifx\csname url\endcsname\relax
  \def\url#1{\texttt{#1}}\fi
\expandafter\ifx\csname urlprefix\endcsname\relax\def\urlprefix{URL }\fi
\providecommand{\bibinfo}[2]{#2}
\providecommand{\eprint}[2][]{\url{#2}}

\bibitem{Sujit2012Delayed}
\bibinfo{author}{Datta, S.~S.} \emph{et~al.}
\newblock \bibinfo{title}{Delayed buckling and guided folding of inhomogeneous
  capsules}.
\newblock \emph{\bibinfo{journal}{Phys. Rev. Lett.}}
  \textbf{\bibinfo{volume}{109}}, \bibinfo{pages}{134302}
  (\bibinfo{year}{2012}).

\bibitem{Paulose2012Fluctuating}
\bibinfo{author}{Paulose, J.}, \bibinfo{author}{Vliegenthart, G.~A.},
  \bibinfo{author}{Gompper, G.} \& \bibinfo{author}{Nelson, D.~R.}
\newblock \bibinfo{title}{Fluctuating shells under pressure}.
\newblock \emph{\bibinfo{journal}{Proceedings of the National Academy of
  Sciences}} \textbf{\bibinfo{volume}{109}}, \bibinfo{pages}{19551--19556}
  (\bibinfo{year}{2012}).

\bibitem{reddy2006theory}
\bibinfo{author}{Reddy, J.~N.}
\newblock \emph{\bibinfo{title}{Theory and analysis of elastic plates and
  shells}} (\bibinfo{publisher}{CRC press}, \bibinfo{year}{2006}).

\bibitem{Landau7}
\bibinfo{author}{Landau, L.} \emph{et~al.}
\newblock \emph{\bibinfo{title}{Theory of Elasticity}}.
\newblock Course of theoretical physics (\bibinfo{publisher}{Elsevier Science},
  \bibinfo{year}{1986}).

\bibitem{audoly2010elasticity}
\bibinfo{author}{Audoly, B.} \& \bibinfo{author}{Pomeau, Y.}
\newblock \emph{\bibinfo{title}{Elasticity and Geometry: From Hair Curls to the
  Non-linear Response of Shells}} (\bibinfo{publisher}{OUP Oxford},
  \bibinfo{year}{2010}).

\bibitem{lidmar2003virus}
\bibinfo{author}{Lidmar, J.}, \bibinfo{author}{Mirny, L.} \&
  \bibinfo{author}{Nelson, D.~R.}
\newblock \bibinfo{title}{Virus shapes and buckling transitions in spherical
  shells}.
\newblock \emph{\bibinfo{journal}{Physical Review E}}
  \textbf{\bibinfo{volume}{68}}, \bibinfo{pages}{051910}
  (\bibinfo{year}{2003}).

\bibitem{Buenemann2007Mechanical}
\bibinfo{author}{Buenemann, M.} \& \bibinfo{author}{Lenz, P.}
\newblock \bibinfo{title}{Mechanical limits of viral capsids}.
\newblock \emph{\bibinfo{journal}{Proceedings of the National Academy of
  Sciences}} \textbf{\bibinfo{volume}{104}}, \bibinfo{pages}{9925--9930}
  (\bibinfo{year}{2007}).

\bibitem{Douglas2001bacteriophage}
\bibinfo{author}{Smith, D.~E.} \emph{et~al.}
\newblock \bibinfo{title}{The bacteriophage $\phi$29 portal motor can package
  dna against a large internal force}.
\newblock \emph{\bibinfo{journal}{Nature}} \textbf{\bibinfo{volume}{413}},
  \bibinfo{pages}{748--752} (\bibinfo{year}{2001}).

\bibitem{phillips2009physical}
\bibinfo{author}{Phillips, R.}, \bibinfo{author}{Kondev, J.} \&
  \bibinfo{author}{Theriot, J.}
\newblock \emph{\bibinfo{title}{Physical Biology of the Cell}}
  (\bibinfo{publisher}{Garland Science}, \bibinfo{year}{2009}).

\bibitem{Seung1988defects}
\bibinfo{author}{Seung, H.~S.} \& \bibinfo{author}{Nelson, D.~R.}
\newblock \bibinfo{title}{Defects in flexible membranes with crystalline
  order}.
\newblock \emph{\bibinfo{journal}{Phys. Rev. A}} \textbf{\bibinfo{volume}{38}},
  \bibinfo{pages}{1005--1018} (\bibinfo{year}{1988}).

\bibitem{Needleman2017}
\bibinfo{author}{Needleman, D.} \& \bibinfo{author}{Dogic, Z.}
\newblock \bibinfo{title}{Active matter at the interface between materials
  science and cell biology}.
\newblock \emph{\bibinfo{journal}{Nature Reviews Materials}}
  \textbf{\bibinfo{volume}{2}}, \bibinfo{pages}{17048} (\bibinfo{year}{2017}).

\bibitem{Scheibner2020}
\bibinfo{author}{Scheibner, C.} \emph{et~al.}
\newblock \bibinfo{title}{Odd elasticity}.
\newblock \emph{\bibinfo{journal}{Nature Physics}}
  \textbf{\bibinfo{volume}{16}}, \bibinfo{pages}{475--480}
  (\bibinfo{year}{2020}).

\bibitem{Salbreux2017}
\bibinfo{author}{Salbreux, G.} \& \bibinfo{author}{J\"ulicher, F.}
\newblock \bibinfo{title}{Mechanics of active surfaces}.
\newblock \emph{\bibinfo{journal}{Phys. Rev. E}} \textbf{\bibinfo{volume}{96}},
  \bibinfo{pages}{032404} (\bibinfo{year}{2017}).

\bibitem{oddreview}
\bibinfo{author}{Fruchart, M.}, \bibinfo{author}{Scheibner, C.} \&
  \bibinfo{author}{Vitelli, V.}
\newblock \bibinfo{title}{Odd viscosity and odd elasticity}
  (\bibinfo{year}{2022}).
\newblock \eprint{Arxiv:2207.00071}.

\bibitem{Truesdell1963Meaning}
\bibinfo{author}{Truesdell, C.~A.}
\newblock \bibinfo{title}{The meaning of betti's reciprocal theorem}.
\newblock \emph{\bibinfo{journal}{Journal of Research of the National Bureau of
  Standards Section B Mathematics and Mathematical Physics}}
  \bibinfo{pages}{85} (\bibinfo{year}{1963}).

\bibitem{Chen2021Realization}
\bibinfo{author}{Chen, Y.}, \bibinfo{author}{Li, X.},
  \bibinfo{author}{Scheibner, C.}, \bibinfo{author}{Vitelli, V.} \&
  \bibinfo{author}{Huang, G.}
\newblock \bibinfo{title}{Realization of active metamaterials with odd
  micropolar elasticity}.
\newblock \emph{\bibinfo{journal}{Nature Communications}}
  \textbf{\bibinfo{volume}{12}}, \bibinfo{pages}{5935} (\bibinfo{year}{2021}).

\bibitem{brandenbourger2021active}
\bibinfo{author}{Brandenbourger, M.}, \bibinfo{author}{Scheibner, C.},
  \bibinfo{author}{Veenstra, J.}, \bibinfo{author}{Vitelli, V.} \&
  \bibinfo{author}{Coulais, C.}
\newblock \bibinfo{title}{Active impact and locomotion in robotic matter with
  nonlinear work cycles}.
\newblock \emph{\bibinfo{journal}{arXiv:2108.08837}}  (\bibinfo{year}{2021}).

\bibitem{bililign2021chiral}
\bibinfo{author}{Bililign, E.~S.} \emph{et~al.}
\newblock \bibinfo{title}{Motile dislocations knead odd crystals into whorls}.
\newblock \emph{\bibinfo{journal}{Nature Physics}}  (\bibinfo{year}{2021}).

\bibitem{tan2021development}
\bibinfo{author}{Tan, T.~H.} \emph{et~al.}
\newblock \bibinfo{title}{Odd dynamics of living chiral crystals}.
\newblock \emph{\bibinfo{journal}{Nature}} \textbf{\bibinfo{volume}{607}},
  \bibinfo{pages}{287--293} (\bibinfo{year}{2022}).

\bibitem{Shankar2022active}
\bibinfo{author}{Shankar, S.} \& \bibinfo{author}{Mahadevan, L.}
\newblock \bibinfo{title}{Active muscular hydraulics}.
\newblock \emph{\bibinfo{journal}{bioRxiv}}  (\bibinfo{year}{2022}).

\bibitem{shankar2020topological}
\bibinfo{author}{Shankar, S.}, \bibinfo{author}{Souslov, A.},
  \bibinfo{author}{Bowick, M.~J.}, \bibinfo{author}{Marchetti, M.~C.} \&
  \bibinfo{author}{Vitelli, V.}
\newblock \bibinfo{title}{Topological active matter}.
\newblock \emph{\bibinfo{journal}{Nature Reviews Physics}}
  \textbf{\bibinfo{volume}{4}}, \bibinfo{pages}{380--398}
  (\bibinfo{year}{2022}).

\bibitem{scheibner2020non}
\bibinfo{author}{Scheibner, C.}, \bibinfo{author}{Irvine, W. T.~M.} \&
  \bibinfo{author}{Vitelli, V.}
\newblock \bibinfo{title}{Non-hermitian band topology and skin modes in active
  elastic media}.
\newblock \emph{\bibinfo{journal}{Phys. Rev. Lett.}}
  \textbf{\bibinfo{volume}{125}}, \bibinfo{pages}{118001}
  (\bibinfo{year}{2020}).

\bibitem{Kane2014}
\bibinfo{author}{Kane, C.~L.} \& \bibinfo{author}{Lubensky, T.~C.}
\newblock \bibinfo{title}{{Topological boundary modes in isostatic lattices}}.
\newblock \emph{\bibinfo{journal}{Nature Physics}}
  \textbf{\bibinfo{volume}{10}}, \bibinfo{pages}{39--45}
  (\bibinfo{year}{2014}).

\bibitem{Bertoldi2017}
\bibinfo{author}{Bertoldi, K.}, \bibinfo{author}{Vitelli, V.},
  \bibinfo{author}{Christensen, J.} \& \bibinfo{author}{van Hecke, M.}
\newblock \bibinfo{title}{Flexible mechanical metamaterials}.
\newblock \emph{\bibinfo{journal}{Nature Reviews Materials}}
  \textbf{\bibinfo{volume}{2}}, \bibinfo{pages}{17066} (\bibinfo{year}{2017}).

\bibitem{huber2016topological}
\bibinfo{author}{Huber, S.~D.}
\newblock \bibinfo{title}{Topological mechanics}.
\newblock \emph{\bibinfo{journal}{Nature Physics}}
  \textbf{\bibinfo{volume}{12}}, \bibinfo{pages}{621--623}
  (\bibinfo{year}{2016}).

\bibitem{Nash2015}
\bibinfo{author}{Nash, L.~M.} \emph{et~al.}
\newblock \bibinfo{title}{{Topological mechanics of gyroscopic metamaterials.}}
\newblock \emph{\bibinfo{journal}{Proc. Natl. Acad. Sci. USA}}
  \textbf{\bibinfo{volume}{112}}, \bibinfo{pages}{14495--500}
  (\bibinfo{year}{2015}).

\bibitem{Wang2015}
\bibinfo{author}{Wang, P.}, \bibinfo{author}{Lu, L.} \&
  \bibinfo{author}{Bertoldi, K.}
\newblock \bibinfo{title}{{Topological Phononic Crystals with One-Way Elastic
  Edge Waves.}}
\newblock \emph{\bibinfo{journal}{Physical review letters}}
  \textbf{\bibinfo{volume}{115}}, \bibinfo{pages}{104302}
  (\bibinfo{year}{2015}).

\bibitem{Susstrunk2015}
\bibinfo{author}{Susstrunk, R.} \& \bibinfo{author}{Huber, S.~D.}
\newblock \bibinfo{title}{{Observation of phononic helical edge states in a
  mechanical topological insulator}}.
\newblock \emph{\bibinfo{journal}{Science}} \textbf{\bibinfo{volume}{349}},
  \bibinfo{pages}{47--50} (\bibinfo{year}{2015}).

\bibitem{Foher2018Spiral}
\bibinfo{author}{Foehr, A.}, \bibinfo{author}{Bilal, O.~R.},
  \bibinfo{author}{Huber, S.~D.} \& \bibinfo{author}{Daraio, C.}
\newblock \bibinfo{title}{Spiral-based phononic plates: From wave beaming to
  topological insulators}.
\newblock \emph{\bibinfo{journal}{Phys. Rev. Lett.}}
  \textbf{\bibinfo{volume}{120}}, \bibinfo{pages}{205501}
  (\bibinfo{year}{2018}).

\bibitem{zhao2020gyroscopic}
\bibinfo{author}{Zhao, Y.}, \bibinfo{author}{Zhou, X.} \&
  \bibinfo{author}{Huang, G.}
\newblock \bibinfo{title}{Non-reciprocal rayleigh waves in elastic gyroscopic
  medium}.
\newblock \emph{\bibinfo{journal}{Journal of the Mechanics and Physics of
  Solids}} \textbf{\bibinfo{volume}{143}}, \bibinfo{pages}{104065}
  (\bibinfo{year}{2020}).

\bibitem{benzoni2020}
\bibinfo{author}{Benzoni, C.}, \bibinfo{author}{Jeevanesan, B.} \&
  \bibinfo{author}{Moroz, S.}
\newblock \bibinfo{title}{Rayleigh edge waves in two-dimensional crystals with
  lorentz forces: From skyrmion crystals to gyroscopic media}.
\newblock \emph{\bibinfo{journal}{Physical Review B}}
  \textbf{\bibinfo{volume}{104}}, \bibinfo{pages}{024435}
  (\bibinfo{year}{2021}).

\bibitem{Sun2020Continuum}
\bibinfo{author}{Sun, K.} \& \bibinfo{author}{Mao, X.}
\newblock \bibinfo{title}{Continuum theory for topological edge soft modes}.
\newblock \emph{\bibinfo{journal}{Phys. Rev. Lett.}}
  \textbf{\bibinfo{volume}{124}}, \bibinfo{pages}{207601}
  (\bibinfo{year}{2020}).

\bibitem{Saremi2020Topological}
\bibinfo{author}{Saremi, A.} \& \bibinfo{author}{Rocklin, Z.}
\newblock \bibinfo{title}{Topological elasticity of flexible structures}.
\newblock \emph{\bibinfo{journal}{Phys. Rev. X}} \textbf{\bibinfo{volume}{10}},
  \bibinfo{pages}{011052} (\bibinfo{year}{2020}).

\bibitem{shankar2021geometric}
\bibinfo{author}{Shankar, S.}, \bibinfo{author}{Bryde, P.} \&
  \bibinfo{author}{Mahadevan, L.}
\newblock \bibinfo{title}{Geometric control of topological dynamics in a
  singing saw} (\bibinfo{year}{2021}).
\newblock \eprint{2108.10875}.

\bibitem{Landau10}
\bibinfo{author}{Pitaevskii, L.} \& \bibinfo{author}{Lifshitz, E.}
\newblock \emph{\bibinfo{title}{Physical Kinetics}}.
\newblock \bibinfo{number}{v. 10} (\bibinfo{publisher}{Elsevier Science},
  \bibinfo{year}{2012}).

\bibitem{braverman2020}
\bibinfo{author}{Braverman, L.}, \bibinfo{author}{Scheibner, C.},
  \bibinfo{author}{VanSaders, B.} \& \bibinfo{author}{Vitelli, V.}
\newblock \bibinfo{title}{Topological defects in solids with odd elasticity}.
\newblock \emph{\bibinfo{journal}{Physical Review Letters}}
  \textbf{\bibinfo{volume}{127}}, \bibinfo{pages}{268001}
  (\bibinfo{year}{2021}).

\bibitem{delplace2017topological}
\bibinfo{author}{Delplace, P.}, \bibinfo{author}{Marston, J.~B.} \&
  \bibinfo{author}{Venaille, A.}
\newblock \bibinfo{title}{Topological origin of equatorial waves}.
\newblock \emph{\bibinfo{journal}{Science}} \textbf{\bibinfo{volume}{358}},
  \bibinfo{pages}{1075--1077} (\bibinfo{year}{2017}).

\bibitem{souslov2019topological}
\bibinfo{author}{Souslov, A.}, \bibinfo{author}{Dasbiswas, K.},
  \bibinfo{author}{Fruchart, M.}, \bibinfo{author}{Vaikuntanathan, S.} \&
  \bibinfo{author}{Vitelli, V.}
\newblock \bibinfo{title}{Topological waves in fluids with odd viscosity}.
\newblock \emph{\bibinfo{journal}{Physical Review Letters}}
  \textbf{\bibinfo{volume}{122}}, \bibinfo{pages}{128001}
  (\bibinfo{year}{2019}).

\bibitem{tauber2019bulk-interface}
\bibinfo{author}{Tauber, C.}, \bibinfo{author}{Delplace, P.} \&
  \bibinfo{author}{Venaille, A.}
\newblock \bibinfo{title}{A bulk-interface correspondence for equatorial
  waves}.
\newblock \emph{\bibinfo{journal}{Journal of Fluid Mechanics}}
  \textbf{\bibinfo{volume}{868}} (\bibinfo{year}{2019}).

\bibitem{tauber2020anomalous}
\bibinfo{author}{Tauber, C.}, \bibinfo{author}{Delplace, P.} \&
  \bibinfo{author}{Venaille, A.}
\newblock \bibinfo{title}{Anomalous bulk-edge correspondence in continuous
  media}.
\newblock \emph{\bibinfo{journal}{Physical Review Research}}
  \textbf{\bibinfo{volume}{2}}, \bibinfo{pages}{013147} (\bibinfo{year}{2020}).

\bibitem{Volovik1988}
\bibinfo{author}{Volovik, G.~E.}
\newblock \bibinfo{title}{An analog of the quantum hall effect in a superfluid
  ${}^{3}$he film}.
\newblock \emph{\bibinfo{journal}{Sov. Phys. JETP}}
  \textbf{\bibinfo{volume}{67}}, \bibinfo{pages}{1804} (\bibinfo{year}{1988}).

\bibitem{So1985}
\bibinfo{author}{So, H.}
\newblock \bibinfo{title}{Induced topological invariants by lattice fermions in
  odd dimensions}.
\newblock \emph{\bibinfo{journal}{Progress of Theoretical Physics}}
  \textbf{\bibinfo{volume}{74}}, \bibinfo{pages}{585--593}
  (\bibinfo{year}{1985}).

\bibitem{Coste1989}
\bibinfo{author}{Coste, A.} \& \bibinfo{author}{Lüscher, M.}
\newblock \bibinfo{title}{Parity anomaly and fermion-boson transmutation in
  3-dimensional lattice {QED}}.
\newblock \emph{\bibinfo{journal}{Nuclear Physics B}}
  \textbf{\bibinfo{volume}{323}}, \bibinfo{pages}{631--659}
  (\bibinfo{year}{1989}).

\bibitem{Bal2019}
\bibinfo{author}{Bal, G.}
\newblock \bibinfo{title}{Continuous bulk and interface description of
  topological insulators}.
\newblock \emph{\bibinfo{journal}{Journal of Mathematical Physics}}
  \textbf{\bibinfo{volume}{60}}, \bibinfo{pages}{081506}
  (\bibinfo{year}{2019}).

\bibitem{Bal2019b}
\bibinfo{author}{Bal, G.}
\newblock \bibinfo{title}{Topological invariants for interface modes}
  (\bibinfo{year}{2019}).
\newblock \eprint{Arxiv:1906.08345v3}.

\bibitem{Avron1998}
\bibinfo{author}{Avron, J.~E.}
\newblock \bibinfo{title}{{Odd Viscosity}}.
\newblock \emph{\bibinfo{journal}{Journal of Statistical Physics}}
  \textbf{\bibinfo{volume}{92}}, \bibinfo{pages}{543--557}
  (\bibinfo{year}{1998}).

\bibitem{Hasan2010}
\bibinfo{author}{Hasan, M.~Z.} \& \bibinfo{author}{Kane, C.~L.}
\newblock \bibinfo{title}{Colloquium: Topological insulators}.
\newblock \emph{\bibinfo{journal}{Rev. Mod. Phys.}}
  \textbf{\bibinfo{volume}{82}}, \bibinfo{pages}{3045--3067}
  (\bibinfo{year}{2010}).

\end{thebibliography}

\clearpage
\onecolumngrid
\appendix

\section{Odd elastic moduli in 3D with cylindrical symmetry}
\label{sec:app-isotropy}

We now construct the most general elastic tensor in a 3D system with cylindrical symmetry i.e. a rank-four tensor which does not change when the coordinate system is rotated around the $z$-axis. 
Since both the strain and the stress are symmetric, we introduce a basis for symmetric rank 2 tensors $\mathcal{G} = \{g^0, \dots, g^5 \}$ by:
\begin{align}
    g^0_{ij}&=\frac{1}{\sqrt 3}
    \begin{pmatrix}
    1   &   0   &   0   \\
    0   &   1   &   0   \\
    0   &   0   &   1   \\
    \end{pmatrix}
    & g^1_{ij}&=\frac{1}{\sqrt 6}
    \begin{pmatrix}
    -1  &   0   &   0   \\
    0   &   -1  &   0   \\
    0   &   0   &   2   \\
    \end{pmatrix} \\
    g^2_{ij}&=\frac{1}{\sqrt 2}
    \begin{pmatrix}
    1   &   0   &   0   \\
    0   &   -1  &   0   \\
    0   &   0   &   0   \\
    \end{pmatrix}
    & g^3_{ij}&=\frac{1}{\sqrt 2}
    \begin{pmatrix}
    0   &   1   &   0   \\
    1   &   0   &   0   \\
    0   &   0   &   0   \\
    \end{pmatrix} \\
    g^4_{ij}&=\frac{1}{\sqrt 2} 
    \begin{pmatrix}
    0   &   0   &   1   \\
    0   &   0   &   0   \\
    1   &   0   &   0   \\
    \end{pmatrix}
    & g^5_{ij}&=\frac{1}{\sqrt 2}
    \begin{pmatrix}
    0   &   0   &   0   \\
    0   &   0   &   1   \\
    0   &   1   &   0   \\
    \end{pmatrix}
\end{align}
Notice that $\mathcal{G}$ has the features that it is easily separated into irreducible representation of SO(3): $g^0$ lies in the trivial representation, and $g^1, \dots , g^5$ (symmetric traceless matrices) lie in a ``spin 2'' irreducible representation. The basis is orthonormal with respect to the trace scalar product, thus an arbitrary symmetric tensor $t_{ij}$ can be decomposed as $t_{ij} = \sum_a t^a g^a_{ij}$ with:
\begin{equation}
    t^a = \sum_{ij} t_{ij} g^a_{ij}.
\end{equation}
The elastic tensor is a linear operator over the space of rank two tensors, and its representative matrix can be calculated by:
\begin{equation}
    C^{ab} = \sum_{ijk\ell} g^a_{ij} C_{ijk\ell} g^b_{k\ell}.
\end{equation}
Hooke's law can thus be expressed as:
\begin{equation}
    \sigma^a = \sum_b C^{ab} u^b.
\end{equation}

Under a rotation $r$ of the reference system, the matrix $g^a$ changes into $g'^a = r g^a r^\mathrm t$ ($\mathrm t$ denotes transposition). Since $g'^a$ is symmetric, too, it can be decomposed on the $\mathcal{G}$ basis, defining the representation of rotations $R$ over the rank-two tensors: $g'^a_{ij} = R^{ab} g^b_{ij}$. Under rotations, the elastic modulus tensor transforms as: $C'^{ab} = R^{am} C^{mn} (R^\mathrm t)^{nb}$. Cylindrical isotropy means that $C$ is invariant under rotations about the $z$-axis, i.e. $C = R C R^\mathrm t$. Equivalently, $C$ commutes with the representation of the generator of rotations about the $z$-axis, $L_z$, which is given by
\begin{equation}
    L_z^{ab} =
    \begin{pmatrix}
    0 & 0 & 0 & 0 & 0 & 0 \\
    0 & 0 & 0 & 0 & 0 & 0 \\
    0 & 0 & 0 & 2 & 0 & 0 \\
    0 & 0 & -2 & 0 & 0 & 0 \\
    0 & 0 & 0 & 0 & 0 & 1 \\
    0 & 0 & 0 & 0 & -1 & 0
    \end{pmatrix}.
\end{equation}
Imposing $[C, L_z] = 0$, one finds that the most general elastic tensor with cylindrical isotropy reads
\begin{equation}
    C^{ab} = 2
    \begin{pmatrix}
        \frac{3}{2} B & D+H & 0 & 0 & 0 & 0 \\
        D- H & \mu_3 & 0 & 0 & 0 & 0 \\
        0 & 0 & \mu_1 & K_1^o & 0 & 0 \\
        0 & 0 & -K_{1}^o & \mu_1 & 0 & 0 \\
        0 & 0 & 0 & 0 & \mu_2 & K_2^o \\
        0 & 0 & 0 & 0 & -K_{2}^o & \mu_2
    \end{pmatrix}.
\end{equation}
Here $B$ is the bulk modulus, which couples spherically symmetric dilations to pressure, $\mu_1, \mu_2, \mu_3$ are shear moduli, which couple shear deformations to the corresponding shear stress, $D$ is a passive mixed modulus that couples passively the cylindrical shear strain to pressure and spherically symmetric dilations to cylindrically symmetric stress. $K^o_1$ is an anti-symmetric coupling between the shears $2$ and $3$ in the $x$-$y$ plane, while $K^o_2$ is an anti-symmetric coupling between shears $4$ and $5$, which could be respectively called $xz$ an $yz$ shears.
In three dimensions, one can show that an isotropic elastic tensor cannot contain odd elastic moduli~\cite{Scheibner2020}. 

\section{Plate constitutive relations}
\label{sec:app-const-rel}
Directly from the parametrization of the displacement field, and thus from the kinematical assumptions on the allowed displacements, we have that $u_{zz} = 0$. Then its conjugate variable, the vertical stress $\sigma_{zz}$, does not affect the dynamics and it is common practice~\cite{reddy2006theory} to set it to zero --- imposing the so called \emph{plane-stress condition}. The constitutive relations of the three-dimensional material with cylindrical isotropy are modified by this assumption and a reduced elastic tensor for the plate, that maps the non-zero strains into the non-zero stresses, is calculated.

We invert the elastic tensor and express it in cartesian coordinates, obtaining a matrix whose rows and columns are both labelled by $xx, yy, xy, xz, yz, zz$. The column of $C^{-1}$ relative to $zz$ spans the strains that produce $\sigma_{zz}$, while the span of the other columns contains the strains that produce plane stresess. We consider only the latter strains and thus delete the column relative to $zz$. We then project the remaining columns  to the space of allowed strains. This is obtained by deleting the row relative to $zz$. The reduced $5 \times 5$ matrix maps the plane stresses into the plane strains. Inverting it, we obtain the elastic tensor for the plate, which maps the allowed strains to the allowed stresses.

More generally, we can consider moduli that are consistent with planar isotropy, but do not arise from the thin limit of a 3D structure. The constitutive relation in this case take the more general form 
\begin{equation}
    \begin{pmatrix}
        N_D \\
        N_{S^1} \\
        N_{S^2} \\
        M_D \\
        M_{S^1} \\
        M_{S^2} \\
        N_{T^x} \\
        N_{T^y}
    \end{pmatrix}
    = 
    \begin{pmatrix}
               C_{00} & 0          & 0         & C_{03}     & 0       & 0      &    0    & 0 \\
               0      & C_{11}     &  C_{12}  &  0           &  C_{14}      &   C_{15}    &    0    &  0\\
               0      & -C_{12}     & C_{11}    & 0          &  -C_{15}      &  C_{14}      &    0     & 0 \\ 
               C_{30} &     0       &    0       &  C_{33} & 0      & 0     &      0  &  0\\
               0      & C_{41}     &  C_{42}     &     0     & C_{44}  & C_{45} &     0   & 0 \\
               0      & -C_{42}    &  C_{41}           &     0     & -C_{45} & C_{44} &     0   & 0 \\
               0      &      0      &   0        &   0        &     0   & 0     &  C_{66} & C_{67} \\
               0      &       0     &  0         &  0         &      0  & 0      &-C_{67}  & C_{66} 
    \end{pmatrix}
    \begin{pmatrix}
        u^0_D \\
        u^0_{S^1} \\
        u^0_{S^2} \\
        u^1_{D} \\
        u^1_{S^1} \\
        u^1_{S^2} \\
        u^0_{T^x} \\
        u^0_{T^y} \\
    \end{pmatrix}
\end{equation}

\section{Virtual work derivation of the equations of motion} \label{sec:app-virtual-work}
In this appendix we calculate the equations describing the evolution of the plate in the bulk using the principle of virtual work. We consider a plate that is infinite along $x$ and $y$ directions, and has finite thickness along $z \in [-h/2, h/2]$. The region occupied by the plate at rest is then $V = \mathbb R^2 \times [-h/2, h/2]$. The displacement field configurations are assumed to vanish at infinity. We consider an external drag force given by $f^{\mathrm{ext}}_i = - \Gamma \dot \xi_i$. 

The virtual work principle (an infinitesimal version of the Hamilton's principle) states that the actual field configuration in its time evolution from $t_i$ to $t_f$ must satisfy
\begin{equation}
\label{eq:virtual-work-thm}
   \int_{t_i}^{t_f} dt ( \delta K - \delta W^{\mathrm{int}} - \delta W^{\mathrm{ext}}) = 0
\end{equation}
where $K$ is the kinetic energy, $\delta W^{\mathrm{int}}$ is the internal infinitesimal work and $\delta W^{\mathrm{ext}}$ is the external infinitesimal work, given by
\begin{equation}
    \begin{aligned}
    K &= \frac{1}{2}\int_V d^3x \, \rho \dot \xi_i \dot \xi_i \\
    \delta W^{\mathrm{int}} &= \int_V d^3x \, \sigma_{ij} \delta u_{ij} \\
    \delta W^{\mathrm{ext}} &= \int_V d^3x \, f^{\mathrm{ext}}_i \delta \xi_i
\end{aligned}
\end{equation}
in which the variations $\delta$ are performed within the allowed displacements \cite{reddy2006theory}. 

In the following, we express the three-dimensional displacement field with respect to the plate fields, then we integrate over $z$, assuming a uniform density. The terms that survive are the ones that contain an even power of $z$. Finally we integrate by parts in order to isolate the variations of the fields. For the kinetic energy, we have:
\begin{align}
    \int_{t_i}^{t_f} dt \, \delta K 
    &= \int_{t_i}^{t_f} dt \int_V  d^3x \, \rho [ (\dot \eta_\alpha + z \dot \phi_\alpha) \delta (\dot \eta_\alpha + z  \dot \phi_\alpha) + \dot w \delta \dot w ] \\
    &= \int_{t_i}^{t_f} dt \int_{\mathbb R^2} d^2x \, \rho [
    h \dot \eta_\alpha \delta \dot \eta_\alpha +  
    h \dot w \delta \dot w + 
    \frac{h^3}{12} \dot \phi_\alpha  \delta \dot \phi_\alpha ] \\
    &= - \int_{t_i}^{t_f} dt \int_{\mathbb R^2} d^2x \, \rho [
    h \ddot \eta_\alpha \delta \eta_\alpha 
    + h\ddot w \delta w 
    + \frac{h^3}{12} \ddot \phi_\alpha \delta\phi_\alpha ].
\end{align} 
For the internal elastic forces, we have:
\begin{align}
    \int_{t_i}^{t_f} dt \, \delta W_{\mathrm{int}} 
    &= \int_{t_i}^{t_f} dt \int_V d^3x \, [
    \sigma_{\alpha\beta} ( \partial_\alpha \delta \eta_\beta + z \partial_\alpha \delta \phi_\beta) + 
    \sigma_{z \alpha} \delta \phi_\alpha + 
    \sigma_{\alpha z} \partial_\alpha \delta w ] \\
    &= \int_{t_i}^{t_f} dt \int_{\mathbb R^2} d^2x \, [
    N_{\alpha\beta} \partial_\alpha \delta\eta_\beta + 
    M_{\alpha\beta} \partial_\alpha \delta \phi_\beta + 
    N_{z\alpha} \delta \phi_\alpha + 
    N_{\alpha z} \partial_\alpha \delta w ] \\
    &= \int_{t_i}^{t_f} dt \int_{\mathbb R^2} d^2x \, [
    -\partial_\alpha N_{\alpha\beta} \delta \eta_\beta 
    - \partial_\alpha M_{\alpha\beta}\delta \phi_\beta 
    + N_{z\alpha}\delta\phi_\alpha - \partial_\alpha N_{\alpha z} \delta w ].
\end{align}
For the external forces, we have:
\begin{align}
    \int_{t_i}^{t_f} dt \, \delta W_{\mathrm{ext}} 
    &= \int_{t_i}^{t_f} dt \int_{\mathbb R^2} d^2x \, \Gamma [h \dot \eta_\alpha \delta \eta_\alpha + h \dot w \delta w + \frac{h^3}{12} \dot \phi_\alpha \delta \phi_\alpha ].
\end{align}
Finally, requiring \cref{eq:virtual-work-thm} to be satisfied for all the variations of the plate's fields, we get 
\begin{subequations}
\begin{align}
    h( \rho \ddot\eta_\alpha + \Gamma \dot \eta_\alpha )  &= \partial_\beta N_{\beta \alpha} \label{eq:eom-eta-raw_SI}\\
    h(\rho  \ddot w  + \Gamma \dot w) &=  \partial_\alpha N_{\alpha z} \label{eq:eom-w-raw_SI} \\
    \frac{h^3}{12}(\rho \ddot \phi_\alpha + \Gamma \dot \phi_\alpha) &= \partial_\beta M_{\beta \alpha} - N_{z \alpha}. \label{eq:eom-phi-raw_SI}
\end{align}
\end{subequations}

Substituting the constitutive relations we get \cref{eq:eom-eta,eq:eom-w,eq:eom-phi}.

\section{Calculation of the first Chern number} \label{sec:app-chern}
Here we show that for a purely active plate, as long as $K^o_2, K^o_1 \neq 0$, momentum space can be compactified to a sphere. Once momentum space is compactified, we may define the first Chern number as a topological invariant for the eigenvector bands, which is later calculated.
The reader is directed to Ref.~\cite{souslov2019topological} and references therein for more details.

We write the purely active dynamical matrix as a linear combination of the matrices
\begin{equation}
    S_x =
    \begin{pmatrix}
    0 & 0 & 0 \\
    0 & 0 & -1 \\
    0 & -1 & 0 
    \end{pmatrix}
    \quad
    S_y = 
    \begin{pmatrix}
    0 & 0 & 1 \\
    0 & 0 & 0 \\
    1 & 0 & 0
    \end{pmatrix}
    \quad
    S_z = 
    \begin{pmatrix}
    0 & -i & 0 \\
    i & 0 & 0 \\
    0 & 0 & 0
    \end{pmatrix}
\end{equation}
with $\mathcal M(\vect q) = i \vec M(\vect q) \cdot \vec S $, and $\vec M(\vect q) = (K^o_2 q_x, K^o_2 q_y, K^o_1 q^2 + K^o_2)$. Since a multiplicative factor does not change the eigenvectors of a matrix, the Hamiltonian $\mathcal N(\mathbf q) = \hat n (\mathbf q) \cdot \vec S$, with $\hat n = \vec M / ||\vec M ||$, has the same eigenvectors as $\mathcal M (\vect q)$ and thus the properties of the bands of eigenvectors of $\mathcal M$ can be studied equivalently on $\mathcal N$. Explicitly, we have
\begin{equation}
    \hat n (\mathbf q) = \frac{1}{\sqrt{q^2 + (1 + K^o_1/K^o_2 q^2)^2}}
    \begin{pmatrix}
        q_x \\
        q_y \\
        (K^o_1/K^o_2) q^2 + 1 \\
    \end{pmatrix}
\end{equation}
which is well defined on the whole momentum space if $K^o_2 \neq 0$. If in addition $K^o_1 \neq 0$ then $\lim_{\vect q \to \infty} \hat n(\vect q)$ does not depend on the chosen direction. 
Identifying all the points at infinity, we compactify the momentum space to a sphere. In this case, the eigenvectors of each band form a vector bundle over a compact manifold and the first Chern number is a well-defined topological invariant. 

\begin{figure}
\begin{center}
\subfloat[][$K^o_1/K^o_2 < 0$.]
    {\includegraphics[width=3.8cm]{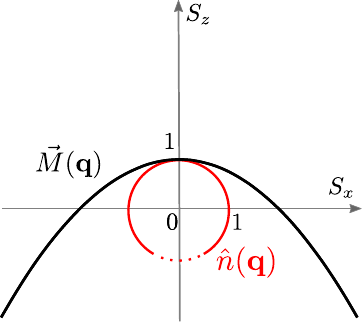}}  
\quad
\subfloat[][$K^o_1/K^o_2 > 0$.]
   {\includegraphics[width=3.8 cm]{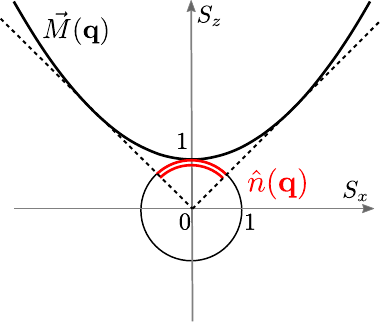}}
\end{center}
\caption{{\bf Visualization of Chern number calculation}~A cross-section of $\vec M (\vect q)$ and $\hat n (\vect q)$ shows the index of the maps.  In red, the projection on the unit sphere. The behaviour of the map $\hat n$ can be understood visualizing the map $\vec M$. (a) $\vec M$ describes a paraboloid that encloses the origin, then $\hat n = \vec M / ||\vec M||$ covers the whole unit sphere. (b) $\vec M$ describes a paraboloid pointing upwards. Then, $\hat n$ covers the same portion of the unit sphere two times, with opposite orientation, giving index zero.}
\label{fig:sphere-cover-gamma}
\end{figure}

The Berry curvature $F$ of the bands of $\mathcal M$ coincides with the Berry curvature of the bands of $\mathcal N$, so we will focus on the latter. We define the following map from three-dimensional unit vectors to $3 \times 3$ complex matrices
\begin{equation}
    \begin{aligned}
    \mathscr S:  \mathbb S^2 &\rightarrow \mathrm{Mat}(3, \mathbb C)\\
     \hat v &\mapsto \hat v \cdot \vec S = v_x S_x + v_y S_y + v_z S_z.
    \end{aligned}
\end{equation}
Then, $\mathcal N$ is the composition of $\hat n$ and $\mathscr S$, i.e. $\mathcal N : \mathbb R^2 \xrightarrow{\hat n} \mathbb S^2 \xrightarrow{\mathscr S} \mathrm{Mat}(3,\mathbb C)$. Each band of eigenvectors of $\mathscr S$ induces a Berry curvature $F^{\mathbb S^2}$ on the sphere. The Berry curvature $F$ induced by $\mathcal N$ is equal to the pull-back through $\hat n$ of $F^{\mathbb S^2}$, i.e. $F = \hat n^* F^{\mathbb S^2}$. The curvature $F^{\mathbb S^2}$ can be calculated explicitly as follows. Let $\hat n = (\sin \theta \cos \phi, \sin \theta \sin \phi, \cos \theta)$, the eigenvectors of the positive and negative bands of $\mathcal N = \hat n \cdot \vec S$ are
\begin{equation}
    \psi_\pm (\hat n) = \frac{1}{\sqrt 2}
    \begin{pmatrix}
    \cos \theta \cos \phi \mp i \sin \theta \\
    \pm i \cos \phi + \cos \theta \sin \phi \\
    -i \sin \theta
    \end{pmatrix}.
\end{equation}
These induce a Berry connection $ A^{\mathbb S^2}_\pm = -i \langle \psi_\pm | d \psi_\pm \rangle = \mp \cos \theta \, d\phi$ and Berry curvature $F^{\mathbb S^2}_\pm = d  A^{\mathbb S^2}_\pm = \pm \sin \theta \, d\theta \wedge d\phi$. We observe that $F_+^{\mathbb S^2}$ is the volume form of the sphere, thus  $\mathcal C_+ = 1/(2\pi) \int_{\mathbb R^2} \hat n^* F^{\mathbb S^2}_+  $ is twice the index (or degree) of the map $\hat n$. The index of a map is an integer that counts the signed number of times that the domain (here the momentum space, compactified to a sphere) wraps on the target space (here the sphere $\mathbb S^2$). The index of $\hat n$ depends on the relative values of $K^o_1$ and $K^o_2$ as follows. When $K^o_1$ and $K^o_2$ have the same sign, the map $\hat n$ does not fully cover the sphere, thereby yielding a vanishing index and $\mathcal C_+ = 0$. 
When $K^o_1$ and $K^o_2$ have opposite sign, the map $\hat n$ covers the sphere once, so $|\mathcal C_+ | = 2$. The sign of the Chern number then depends on the orientation of the covering: 
if $K^o_2 > 0$, $\hat n(\vect 0) = (0,0,1)^\mathrm t$, so the sphere is covered from the top and thus $\mathcal C_+ = 2$; if $K^o_2 < 0$, the sphere is covered from the bottom, leading to $\mathcal C_- = -2$. The result is summarized in \cref{eq:chern-result}. A pictorial representation is given in \cref{fig:sphere-cover-gamma}.

\section{Mapping between odd-elastic plates and odd-viscous fluids}
\label{app:mapping}

We assume that the dynamics of the active plate is overdamped, which means that inertial terms in the equations ($\rho \omega^2$ in Eq.~\eqref{eq:eigenvalue-problem}) can be neglected over the drag term ($i \Gamma \omega$ in Eq.~\eqref{eq:eigenvalue-problem}). We can compare the resulting equations to those of a fluid with odd viscosity exposed to an external magnetic field~\cite{souslov2019topological}. For the fluid with odd viscsoity, the physical fields are the density $\rho$ and the velocity $\vect v$. The parameters are the average density $\rho_0$, the speed of sound $c$, a typical frequency $\omega_B$ analogous to the cyclotron frequency, the ordinary viscosity $\nu$ and odd viscosity $\nu^0$. The non-dimensional linearized Navier-Stokes equations are (\cite{souslov2019topological}, Appendix VII)
\begin{equation}
    \begin{aligned}
    \partial_{t} \mathbf{v} &= -\mathrm{Ma}^{-2} \nabla (\rho / \rho_0) +\mathrm{Ro}^{-1} \mathbf{v}^{*} 
    \\ 
    & \qquad + \mathrm{Re}^{-1} \nabla^{2} \mathbf{v} + \mathrm{Re}_{\mathrm{odd}}^{-1} \nabla^{2} \mathbf{v}^{*} \\
    \partial_t  (\rho/\rho_0) &= -  \nabla \cdot \vect v
    \end{aligned}
\end{equation}
where $\vect v^* \equiv (v_y, -v_x)$ is the velocity rotated by $90^{\circ}$ and the Mach, Rossby, Reynolds and odd Reynolds dimensionless numbers are respectively defined as
\begin{align}
    \mathrm{Ma} &= \frac{U}{c} & \mathrm{Ro} &= \frac{U}{L \omega_B} & \mathrm{Re} &= \frac{UL}{\nu} & \mathrm{Re}_\mathrm{odd} &= \frac{UL}{\nu^o}.
\end{align}

The non-dimensionalized equations of an odd-elastic plate with $\tilde B = \mu_2 = 0$ can be obtained defining a time scale $T$ (so that  $\partial_t \to T^{-1} \partial_t$) and read
\begin{equation}
    \begin{aligned}
    \partial_t \vect \phi &= - \bar K^o_2 \nabla^* w - \bar K^o_2 \vect \phi^* + \bar \mu_1 \nabla^2 \vect \phi + \bar K^o_1 \nabla^2 \vect \phi^* \\
    \partial_t w &= \bar K^o_2 \nabla \cdot \vect \phi^*
    \end{aligned}
\end{equation}
with
\begin{align}
    \bar \mu_1 &= \frac{12 \mu_1 T}{h^2 \Gamma} & \bar K^o_1 &= \frac{12 K^o_1 T}{h^2 \Gamma} & \bar K^o_2 &= \frac{12 K^o_2 T}{h^2 \Gamma}.
\end{align}
and $\nabla^* \equiv (\partial_y, -\partial_x)$. Then the mapping from the odd-elastic plate to the odd-viscous fluid is obtained through the identification
\begin{equation}
    \begin{aligned}
    \phi_\alpha &= \epsilon_{\alpha \beta} v_\beta & w &= \rho / \rho_0 \\
    \mathrm{Ma}^{-2} &= \bar K^o_2  & \mathrm{Ro}^{-1} &= - \bar K^o_2 \\ 
    \mathrm{Re}^{-1} &= \bar \mu_1 & \mathrm{Re}_{\mathrm{odd}}^{-1} &= \bar K^o_1 \\ 
    1 &= \bar K^o_2.
\end{aligned}  
\end{equation}
In the mapping, the two scalar fields are identified and the two vector fields are identified (after a rotation by $90^\circ$). $K^o_1$ is related to the odd Reynolds number, as they refer to the same tensor component, respectively for the elastic tensor and for the viscosity tensor. $K^o_2$ is related to $\omega_B$. In fact, $K^o_2$ makes the vector $\phi$ rotate, as $\omega_B$ makes the velocity rotate. The passive shear modulus $\mu_1$ acts as an ordinary shear viscosity term. Finally, the requirement $1 = \bar K^o_2$ fixes the time scale of the plate to $T = h^2 \Gamma / (12 K^o_2)$. 

\end{document}